\documentclass[reprint,aps,prd,superscriptaddress,showkeys,showpacs]{revtex4-1}
\usepackage{epsfig,amsmath,natbib}

\usepackage{aas_macros}
\usepackage{amssymb}
\usepackage{amsmath}
\usepackage{dsfont}
\usepackage{hyperref}
\usepackage{color}
\usepackage{pbox}
\usepackage{booktabs}
\usepackage[dvipsnames]{xcolor}

\hypersetup{
	colorlinks=false,
	citecolor=green
}

\newcommand{\bb}[1]{\mathbf{#1}}
\newcommand{\bbh}[1]{\mathbf{\hat{#1}}}
\newcommand{\h}[1]{\hat{#1}}

\newcommand{\ttt}[1]{\texttt{#1}}

\newcommand\lsim{\mathrel{\rlap{\lower4pt\hbox{\hskip1pt$\sim$}}
        \raise1pt\hbox{$<$}}}
\newcommand\gsim{\mathrel{\rlap{\lower4pt\hbox{\hskip1pt$\sim$}}
        \raise1pt\hbox{$>$}}}

\begin{document}

\title{Cosmology with photometric weak lensing surveys: constraints with redshift tomography of convergence peaks and moments}

\author{Andrea Petri}
\email{apetri@phys.columbia.edu}
\affiliation{Department of Physics, Columbia University, New York, NY 10027, USA}
\affiliation{Physics Department, Brookhaven National Laboratory, Upton, NY 11973, USA}

\author{Morgan May}
\affiliation{Physics Department, Brookhaven National Laboratory, Upton, NY 11973, USA}

\author{Zolt\'an Haiman}
\affiliation{Department of Astronomy, Columbia University, New York, NY 10027, USA}

\date{\today}

\label{firstpage}

\begin{abstract}
Weak gravitational lensing is becoming a mature technique for
constraining cosmological parameters, and future surveys will be able
to constrain the dark energy equation of state $w$. When analyzing
galaxy surveys, redshift information has proven to be a valuable
addition to angular shear correlations. We forecast parameter
constraints on the triplet $(\Omega_m,w,\sigma_8)$ for an LSST-like
photometric galaxy survey, using tomography of the shear-shear power
spectrum, convergence peak counts and higher convergence moments. We
find that redshift tomography with the power spectrum reduces the area
of the $1\sigma$ confidence interval in $(\Omega_m,w)$ space by a
factor of 8 with respect to the case of the single highest redshift
bin. We also find that adding non-Gaussian information from the peak
counts and higher-order moments of the convergence field and its
spatial derivatives further reduces the constrained area in
$(\Omega_m,w)$ by a factor of 3 and 4, respectively. When we add
cosmic microwave background parameter priors from Planck to our
analysis, tomography improves power spectrum constraints by a factor
of 3. Adding moments yields an improvement by an additional factor of
2, and adding both moments and peaks improves by almost a factor of 3,
over power spectrum tomography alone. We evaluate the effect of
uncorrected systematic photometric redshift errors on the parameter
constraints. We find that different statistics lead to different bias
directions in parameter space, suggesting the possibility of
eliminating this bias via self-calibration. 
\end{abstract}

\keywords{Weak gravitational lensing --- Simulations --- Systematic effects: photometry --- Methods: numerical, statistical}
\pacs{98.80.-k, 95.36.+x, 95.30.Sf, 98.62.Sb}

\maketitle


\section{Introduction}
Weak gravitational lensing is a promising technique to probe the large scale structure of the universe in which the tracers are intrinsically unbiased \citep{wlreview}. This technique has the potential of significantly improving the constraints on the dark energy equation of state parameter $w$ because it is most sensitive to the matter density fluctuations at the non--linear stage. Cosmology inferences from weak lensing observations have been produced for past (CFHTLenS \citep{cfht1}, COSMOS \citep{cosmos}) and current (DES \citep{DES}) surveys, and are being planned for future experiments as well (e.g. LSST \citep{LSST}, WFIRST \citep{WFIRST}, Euclid \citep{Euclid}). Because of the non-linear nature of the density fluctuations probed by weak lensing, cosmological information might leak from quadratic statistics (such as two-point functions and power spectra) into more complicated non-Gaussian statistics, for which forward modeling requires numerical simulations of cosmic shear fields. 

Several different examples of these non-Gaussian statistics, and their cosmological information content, have been studied in the past as well (see \citep{MinkJan,PeaksJan,NG-Marian,NG-Jain1,NG-Jain2,NG-Jain3,NG-Refregier,NG-Dietrich} for a non-comprehensive list). The constraining power of weak lensing power spectra with the addition of redshift tomography information have been extensively investigated in the literature (see e.g. \citep{SongKnox,FangHaiman07,Huterer2006}). In this work we concentrate on the constraining power of a subset of non-Gaussian statistics, combined with redshift tomography in an LSST-like survey. \citep{MartinetPeaksTomo} investigated the cosmological constraining power of shear peaks tomography. Previous work on redshift tomography with weak lensing Minkowski functionals is also present in the literature \citep{MinkJan}. 

Tomography relies on assigning accurate redshifts to galaxies. We therefore also investigate the effects of uncorrected photometric redshift systematics on parameter constraints when using redshift tomography. This work is organized as follows: in \S~\ref{sec:methods} we outline the shear simulations we use in this work, followed by descriptions of the convergence reconstruction procedure, forward modeling of galaxy shape and photometric redshift systematics, and the parameter-inference techniques we used to forecast constraints on cosmology. In \S~\ref{sec:results} we present our main results, which we discuss in \S~\ref{sec:discussion}. In \S~\ref{sec:conclusions} we present our conclusions as well as prospects for future work.  


\section{Methods}
\label{sec:methods}


\subsection{Cosmic shear simulations}
\label{sec:shearsim}
We review the procedure used for generating simulated shear catalogs. We consider a fiducial flat $\Lambda$CDM universe with parameters $(h,\Omega_m,\Omega_\Lambda,\Omega_b,w,\sigma_8,n_s)=(0.72,0.26,0.74,0.046,-1,0.8,0.96)$ \citep{WMAP9,PlanckCosmo}. We examine different variations of the $\bb{p}=(\Omega_m,w,\sigma_8)$ triplet and run one $N$--body simulation for each choice of $\bb{p}$, using the public code \ttt{Gadget2} \citep{Gadget2}. The simulations have a comoving box size of $L_b=260\,{\rm Mpc}/h$ and contain $512^3$ dark matter particles, which correspond to a mass resolution of $M_p\approx 10^{10}M_{\rm sun}$ per particle. 

The largest mode observed in our $N$--body simulations corresponds to a wavenumber of $k_b\approx1/L_b\approx 0.004 h{\rm Mpc}^{-1}$. For the sake of recovering cosmological information from WL, this limitation does not create a concern, as several authors (see \citep{FangHaiman07} for example) have shown that modes above $L_b$ contribute very little to parameter constraints. 
Moreover, the purpose of this work is to estimate the parameter constraints achievable in a weak lensing analysis incorporating tomography, not to produce simulations accurate enough for analyzing the data set that will be available from LSST and other surveys a decade hence. To analyze the datasets that these surveys will produce, mode couplings between large and small scales, which can cause effects such as super sample covariance \citep{Sato12,SSC1,SSC2}, will need to be included. Baryonic effects will need to be included as well. Larger and more accurate $N$--body simulation techniques are currently under development in the community for this purpose \citep{Qcontinuum,HACC}.

The three dimensional outputs of the $N$--body simulations are sliced in sequences of two dimensional lenses $120 \,{\rm Mpc}$ thick, which are lined up perpendicular to the line of sight between the observer on Earth and a source at redshift $z_s$. We make use of the multi--lens--plane algorithm \citep{RayTracingJain,RayTracingHartlap} to trace the deflections of light rays originating at $z=0$ through the system of lenses out to redshift $z$. To accomplish this task, we make use of the \ttt{LensTools} \citep{LensTools-ASCL,LensTools-paper} implementation of the multi--lens--plane algorithm. An observed galaxy position $\pmb{\theta}$ on the sky today corresponds to a real galaxy angular position $\pmb{\beta}(\pmb{\theta},z_s)$, which can be calculated using the \ttt{LensTools} pipeline by solving the ordinary differential lens equations up to redshift $z_s$. The Jacobian of $\pmb{\beta}(\pmb{\theta},z_s)$ is a $2\times 2$ matrix that contains information about the cosmic shear field at $\pmb{\theta}$ integrated along the line of sight. 
\begin{equation}
\label{meth:sheareqn}
\frac{\partial\beta_i(\pmb{\theta},z_s)}{\partial \theta_j} = 
\begin{pmatrix}
1-\kappa(\pmb{\theta})-\gamma^1(\pmb{\theta}) & -\gamma^2(\pmb{\theta}) \\
-\gamma^2(\pmb{\theta}) & 1-\kappa(\pmb{\theta})+\gamma^1(\pmb{\theta})\\
\end{pmatrix}
\end{equation}  
The quantities that appear in equation (\ref{meth:sheareqn}) are the convergence $\kappa$, which is the source magnification due to lensing, and the cosmic shear $\pmb{\gamma}$, which is a measurement of the source ellipticity due to lensing from large scale structure, assuming the non-lensed shape is a circle. 

\begin{figure*}
\includegraphics[scale=0.35]{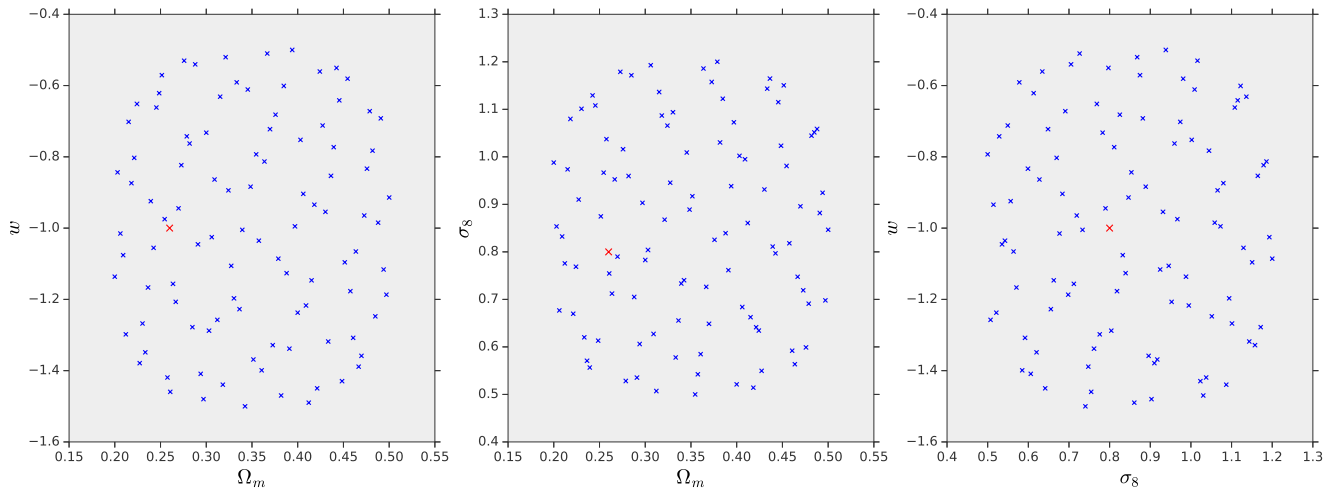}
\caption{Parameter space sampling chosen for our simulations. We show the sampling in the $(\Omega_m,w)$ (left), $(\Omega_m,\sigma_8)$ (center) and $(\sigma_8,w)$ (right) projections of the parameter space. The fiducial cosmology has been marked in red.}
\label{fig:simdesign}
\end{figure*}

We simulate $N_g = 10^6$ random galaxy positions $\{\pmb{\theta}_g\}$ distributed uniformly in a field of view of size $\theta_{\rm FOV}^2=(3.5{\rm deg})^2$, which correspond to a galaxy surface density of $n_g=22\,\rm{arcmin}^{-2}$. The galaxies have a distribution in redshift which mimics the one expected in the LSST survey,
\begin{equation}
\label{meth:galdistr}
n(z) = n_0\left(\frac{z}{z_0}\right)^2\exp\left(-\frac{z}{z_0}\right) ,
\end{equation}  
with $z_0=0.3$ and $n_0$ a normalization constant fixed so that $n(z)$ integrates to the total number of galaxies $N_g$. The galaxies have a maximum redshift $z_{\rm max}=3$. For each galaxy, we compute the cosmic shear at $\pmb{\theta}_g$ using equation (\ref{meth:sheareqn}), producing a shear catalog $\{\pmb{\gamma}_g\}$. Different random realizations of a shear catalog $\{\pmb{\gamma}_g\}_r$ can be obtained rotating and periodically shifting the large scale structure in the $N$--body snapshots according to the procedure explained in \citep{PetriVariance}. We produce $N_r=16000$ pseudo--independent realizations of the shear catalog $\{\pmb{\gamma}_g\}$. These shear realizations all together cover 10 times the total survey area of LSST.
We repeat the above procedure for $P=100$ different combinations of the parameter triplet $\bb{p}$, sampled according to a Latin hypercube scheme. The sampling procedure is the same as described in \citep{CFHTMink,CFHTPeaks}. The parameter space sampling we adopted for our simulations is shown in Figure \ref{fig:simdesign}.

For each of the parameter choices in Figure \ref{fig:simdesign}, the $N$--body initial conditions are generated using the same random seed. In addition to these simulations, we produce simulated shear catalogs for a fiducial $\Lambda$CDM universe with $\bb{p}_0=(0.26,-1,0.8)$. In this case the randomization procedure is based on 5 independent $N$--body simulations, and the same number $N_r=16000$ of pseudo--independent catalog realizations is produced. This additional simulation set serves two purposes: it provides an independent dataset from which to measure covariance matrices, and it provides a way to construct simulated observations that are independent of the simulations on which the cosmological forward model is trained. For the fiducial dataset we chose to base the shear randomization procedure on 5 independent $N$--body simulations to ensure the independence of the $N_r$ realizations for the purpose of estimating covariance matrices. Ref. \citep{PetriVariance} recently showed that, even with only one $N$--body simulation a few $10^4$ independent realizations can be produced.   


\subsection{Forward modeling of systematics}
We next give an overview of the shear systematics included in this work. 

The measured galaxy ellipticity $\pmb{\epsilon}$ is an estimate of the cosmic shear $\pmb{\gamma}$ due to large scale structure if the non--lensed galaxy shape is a circle. Because the galaxies have intrinsic noncircular shapes, the measured galaxy ellipticity $\pmb{\epsilon}_{\rm m}$ is the sum of a cosmic shear term and the intrinsic ellipticity (galaxy shape noise) \citep{wlreview}
\begin{equation}
\pmb{\epsilon}_{\rm m} = \pmb{\gamma} + \pmb{\epsilon}_{\rm n}
\end{equation} 
where $\pmb{\epsilon}_{\rm n}$ is a random Gaussian variable with zero mean and redshift dependent variance $\sigma_n(z_s)=0.15+0.035z_s$. This is equivalent to saying that the cosmic shear inferred from ellipticity observations $\pmb{\gamma}_{\rm m}$ can be written as the sum of the true cosmic shear plus a noise term $\pmb{\gamma}_{\rm n}$ with the same statistical properties as $\pmb{\epsilon}_{\rm n}$. 
We add independent random realizations of the shape noise $\pmb{\gamma}_{\rm n}$ to each of the $N_r$ shear catalogs. Each shape noise realization is generated with a different random seed. The same random seeds are used to generate shape noise catalogs across simulations with different cosmological parameters $\{\bb{p}_i\}$.

In addition to shape noise contributions to the observed galaxy ellipticity, we consider photometric redshift errors as an additional contamination in the simulated catalogs. In photometric surveys such as LSST, the source redshift $z_s$ is estimated measuring the source luminosity in a small finite set of optical frequency bands. Using this compressed luminosity information rather than the full spectrum introduces biases in redshift estimation. Forward modeling of the cosmic shear using the procedure described in \S~\ref{sec:shearsim}, as well as the shape noise contributions, assume a correct redshift distributions $n(z)$. An incorrect binning of observed galaxy redshifts according to the measured photometric distribution $n_p(z_p)$ can propagate the redshift measurement errors all the way to cosmological parameter constraints if the latter take advantage of redshift tomography. One of the goals of this work is to evaluate the size of this effect, assuming photometric redshift errors (photo-$z$) are left uncorrected. 

The study of photometric redshift errors is an active subject of research, and includes investigation of techniques such as spectroscopic calibration, catastrophic errors and cross-correlation techniques that we do not explore in this work (see for example \citep{HuTomo,LSSTSciBook} for a more thorough discussion). We model the effect of photo-$z$ errors as a constant bias term $b_{\rm ph}(z_s)$ plus a random Gaussian component with variance $\sigma_{\rm ph}(z_s)$,
\begin{equation}
\label{meth:photoz-correction}
z_p(z_s) = z_s + b_{\rm ph}(z_s) + \sigma_{\rm ph}(z_s)\mathcal{N}(0,1),   
\end{equation}
where $\mathcal{N}(0,1)$ is the standard normal distribution. We bin the $N_g$ galaxies in our simulated catalogs into 5 redshift bins $\bar{z}_b$, $b=1...5$. Several models have been proposed in the literature for the photometric bias $b_{\rm ph}(z_s)$ (see for example \citep{Huterer2006}) and variance $\sigma_{\rm ph}(z_s)$ (see for example \citep{LSSTSciBook}). We chose the photo-$z$ bias and variance functions in equation (\ref{meth:photoz-correction}) to be the science requirements contained in the LSST Science Book \citep{LSSTSciBook}, namely $b(z_s)=0.003(1+z_s)$ and $\sigma(z_s)=0.02(1+z_s)$. 

\begin{figure}
\includegraphics[scale=0.3]{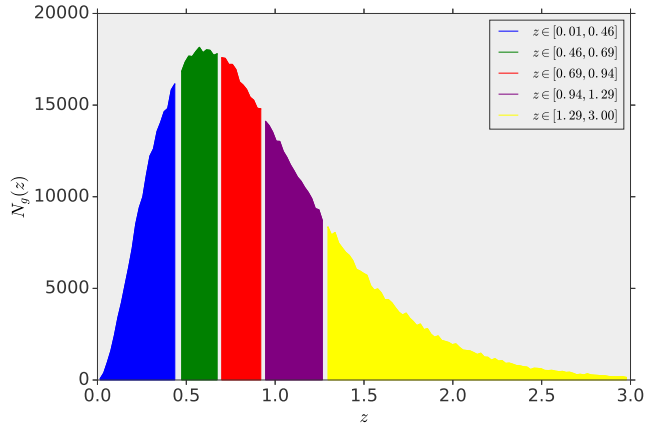}
\caption{Galaxy distribution assumed throughout this work (see equation (\ref{meth:galdistr})), along with the choice of the redshift bins $\bar{z}_b$. Our galaxy sample consists of $N_g=10^6$ galaxies. The figure shows the number of galaxies $N_g(z)$ at each redshift $z$.}
\label{fig:galdistr}
\end{figure}

We generate simulated observations by applying an independent random realization of the photo-$z$ correction (\ref{meth:photoz-correction}) to each catalog realization in the fiducial cosmology $\bb{p}_0$ and by re--binning the galaxies according to their photometric redshifts $z_p$. In the remainder of the paper we use the following notation: we indicate a shear realization $r$ in cosmology $\bb{p}$ with shape noise added as $\h{\pmb{\gamma}}_r(\pmb{\theta}_g,z_g;\bb{p})$, and we indicate a simulated observation in the fiducial cosmology as $\h{\pmb{\gamma}}_{\rm obs}(\pmb{\theta}_g,z_g)$.           


\subsection{Convergence reconstruction}
In this section we describe the procedure we use to construct convergence maps $\kappa$ from the simulated shear catalogs $\pmb{\gamma}$. We consider a two dimensional square pixel grid of area $\theta_{\rm FOV}^2$ and with 512 pixel per side. This correspond to a linear pixel resolution of $0.5 \rm{arcmin}$. We assign a shear value $\pmb{\gamma}(\pmb{\theta}_p,\bar{z}_b)$ to each pixel $\pmb{\theta}_p$ according to the following procedure
\begin{equation}
\pmb{\gamma}(\pmb{\theta}_p,\bar{z}_b) = \frac{\sum_{g=1}^{N_g}\pmb{\gamma}(\pmb{\theta}_g,z_g)W(\pmb{\theta}_g,\pmb{\theta}_p;z_g,\bar{z}_b)}{\sum_{g=1}^{N_g}W(\pmb{\theta}_g,\pmb{\theta}_p;z_g,\bar{z}_b)}
\end{equation}   
We chose a top--hat window function
\begin{equation}
W(\pmb{\theta}_g,\pmb{\theta}_p;z_g,\bar{z}_b) = 
\begin{cases}
1 \,\,\,\,{\rm if}\,\,\,\, \pmb{\theta}_g\in\pmb{\theta}_p,z_g\in\bar{z}_b \\
0 \,\,\,\,{\rm otherwise}
\end{cases}
\end{equation} 
The convergence $\kappa(\pmb{\theta}_p)$ can be reconstructed from the $E$--mode of the shear field, which is evaluated from the Fourier transform of the pixelized shear $\pmb{\gamma}(\pmb{\theta}_p,\bar{z}_b)$
\begin{equation}
\label{meth:psdefinition}
\tilde\kappa(\pmb{\ell},\bar{z}_b) = \left(\frac{\tilde{\gamma}^1(\pmb{\ell},\bar{z}_b)(\ell_x^2-\ell_y^2)+2\ell_x\ell_y\tilde{\gamma}^2(\pmb{\ell},\bar{z}_b)}{\ell_x^2+\ell_y^2}\right) e^{-\frac{\ell^2\theta_G^2}{2}}
\end{equation}
We chose the Gaussian filter smoothing scale $\theta_G=0.5\,{\rm arcmin}$ to correspond to the linear pixel resolution. Inverting the Fourier transform yields the pixelized map $\kappa(\pmb{\theta}_p,\bar{z}_b)$. We apply this procedure to both the shear realizations $\h{\pmb{\gamma}}_r(\pmb{\theta}_g,z_g;\bb{p})$ and the simulated observations $\h{\pmb{\gamma}}_{\rm obs}(\pmb{\theta}_g,z_g)$, yielding convergence realizations $\h{\kappa}_r(\pmb{\theta}_p,\bar{z}_b;\bb{p})$ and simulated convergence observations $\h{\kappa}_{\rm obs}(\pmb{\theta}_p,\bar{z}_b)$. 

We measure a variety of summary statistics from the pixelized convergence maps, which will then be used to forecast parameter constraints and biases. We consider three kinds of summary statistics, namely the tomographic power spectrum $P^{\kappa\kappa}(\ell,\bar{z}_b,\bar{z}_{b'})$, the tomographic peak counts $n_{\rm pk}(\nu,\bar{z}_b)$ and a set of moments $\pmb{\mu}(\bar{z}_b)$. The tomographic power spectrum is defined as 
\begin{equation}
\langle\tilde{\kappa}(\pmb{\ell},\bar{z}_b)\tilde{\kappa}(\pmb{\ell}',\bar{z}_{b'})\rangle = (2\pi)^2\delta_D(\pmb{\ell}+\pmb{\ell}')P^{\kappa\kappa}(\ell,\bar{z}_b,\bar{z}_{b'})
\end{equation}
Because the $\kappa$ field is statistically isotropic, the expectation value $\langle\rangle$, for each realization $r$, is taken over all modes $\pmb{\ell}$ with the same magnitude $\ell=\vert\pmb{\ell}\vert$. Given the fact that our simulation box is small, and we are ignoring non--linear couplings between large and small scale modes, we are likely underestimating the $\kappa$ power spectrum when performing ensemble averages based on a single $N$--body box. \citep{NbodySample} estimated the effect of a varying box size on the 3D matter power spectrum, for boxes up to 512Mpc$/h$ in size and found the variations to be small compared to their sample variance, on spatial wavenumbers up to $k\sim 0.3h{\rm Mpc}^{-1}$. 

The peak count statistic $n_{\rm pk}(\nu,\bar{z}_b)$ is defined as the number of the local maxima of a certain height $\kappa_{\rm max}=\nu\sigma_0$, where $\sigma_0$ is the standard deviation over all pixels. The set of nine moments $\pmb{\mu}(\bar{z}_b)$ is defined as follows (see \citep{Matsubara10,Munshi12,MinkPetri}): 
\begin{equation}
\label{meth:momdef}
\begin{matrix}
\pmb{\mu} = (\pmb{\mu}_2,\pmb{\mu}_3,\pmb{\mu}_4) \\ \\
\pmb{\mu}_2 = (\langle\kappa^2\rangle,\langle\vert\nabla\kappa\vert^2\rangle) \\ \\
\pmb{\mu}_3 = (\langle\kappa^3\rangle,\langle\kappa\vert\nabla\kappa\vert^2\rangle,\langle\kappa^2\nabla^2\kappa\rangle) \\ \\
\pmb{\mu}_4 = (\langle\kappa^4\rangle_c,\langle\kappa^2\vert\nabla\kappa\vert^2\rangle_c,\langle\kappa^3\nabla^2\kappa\rangle_c,\langle\vert\nabla\kappa\vert^4\rangle_c) .
\end{matrix}
\end{equation}
In equation (\ref{meth:momdef}) the gradients $\nabla$ are evaluated using finite differences between $\kappa$ values at neighboring pixels and the expectation values $\langle\rangle$ for each realization $r$ are taken over the $512^2$ pixels in the map. The subscript $c$ indicates that we consider only the connected parts of the quartic $\kappa$ moments. In the definition of the peak counts and convergence moments we omitted the redshift index $\bar{z}_b$ for notational simplicity. In the next section, we describe the statistical methods we use to infer cosmological parameter estimates $\bbh{p}$ from simulated observations $\h{\kappa}_{\rm obs}(\pmb{\theta}_p,\bar{z}_b)$ using the summary statistics $P^{\kappa\kappa}(\ell,\bar{z}_b,\bar{z}_{b'})$, $n_{\rm pk}(\nu,\bar{z}_b)$ and $\pmb{\mu}(\bar{z}_b)$. Concerns might arise on the accuracy with which our simulations measure the summary statistics mentioned above, given the small box size and the fact that we recycle a single $N$--body box for building our simulated sample. \citep{PetriVariance} studied the dependence of the power spectrum and peak counts sample means as a function of the number of independent $N$--body boxes and found that the variations are less than 10\% in most cases, except for the small scale power spectrum and the highest $\kappa$ peaks, for which the variations are less than 20\%. 


\subsection{Parameter inference}
We adopt a Bayesian framework to forecast parameter constraints. We indicate as $\bb{d}$ a summary statistic vector (which can be any of $P^{\kappa\kappa},n_{\rm pk},\pmb{\mu}$ or a combination of these). We label $\bb{d}(\bb{p})$ the sample mean of $\bb{d}$ over the $N_r=16000$ simulated realizations in cosmology $\bb{p}$ and we label $\bbh{d}_r$ the summary statistic measured in realization $r$ of the fiducial cosmology $\bb{p}_0$. Both $\bb{d}(\bb{p}),\bbh{d}_r$ are measured taking galaxy shape noise into account. We further label $\bbh{d}_{\rm obs}$ the summary statistic measured in a simulated observation in which $\kappa$ has been measured taking photo-$z$ errors into account. $\bbh{d}_{\rm obs}$ is measured averaging a random sample of $N_{\rm FOV}=1600$ realizations of the fiducial cosmology with photo-$z$ errors added. This number has been chosen to mimic the survey area of LSST $\Omega_{\rm LSST}=N_{\rm FOV}\theta_{\rm FOV}^2$. Assuming no prior knowledge of the parameters $\bb{p}$, we can write the parameter likelihood $\mathcal{L}$ given the observation $\bbh{d}_{\rm obs}$ using Bayes' theorem
\begin{equation}
\label{meth:paramlikelihood}
-2\log\mathcal{L}(\bb{p}\vert \bbh{d}_{\rm obs}) = (\bbh{d}_{\rm obs} - \bb{d}(\bb{p}))^T\bb{C}^{-1}(\bbh{d}_{\rm obs} - \bb{d}(\bb{p}))
\end{equation}  
The parameter likelihood (\ref{meth:paramlikelihood}) can be evaluated at every point $\bb{p}$ in parameter space by interpolating $\bb{d}(\bb{p})$ between simulation points $\{\bb{p}_i\}$ using a Radial Basis Function (RBF) interpolation (see \citep{CFHTMink,LensTools-paper}). $\bb{C}$ is the $\bb{d}$ covariance matrix and is assumed to be $\bb{p}$--independent. In practice we replace $\bb{C}$ with its estimated value $\bbh{C}$ from $N_r=16000$ realizations of the summary statistics $\bbh{d}_r$ in the fiducial cosmology $\bb{p}_0$ without photo-$z$ errors
\begin{equation}
\bbh{d}_{\rm mean} = \frac{1}{N_r}\sum_{r=1}^{N_r}\bbh{d}_r ,
\end{equation}
\begin{equation}
\bbh{C} = \frac{1}{N_r-1}\sum_{r=1}^{N_r} (\bbh{d}_r-\bbh{d}_{\rm mean})(\bbh{d}_r-\bbh{d}_{\rm mean})^T .
\end{equation} 
Cosmological parameter values $\bbh{p}$ can be inferred from equation (\ref{meth:paramlikelihood}) by looking at the location at which the likelihood $\mathcal{L}(\bb{p}\vert\bbh{d}_{\rm obs})$ is maximum. Parameter errors $\Delta\bbh{p}$ can be inferred from the likelihood confidence contours. Estimates of $\bbh{p},\Delta \bbh{p}$ can be obtained by approximating the model statistic $\bb{d}(\bb{p})$ dependency on parameters as linear in $\bb{p}$, provided $\bb{p}$ is not too far from the fiducial model $\bb{p}_0$
\begin{equation}
\label{meth:linapprox}
\bb{d}(\bb{p}) = \bb{d}(\bb{p}_0) + \bb{M}(\bb{p}-\bb{p}_0) + O(\vert\bb{p}-\bb{p}_0\vert^2)
\end{equation} 
where we defined $(\bb{M})_{i\alpha}=(\partial d_i(\bb{p})/\partial p_\alpha)_{\bb{p}_0}$ as the first derivative of the statistic $d_i(\bb{p})$ with respect to cosmology. We evaluate $\bb{M}$ with finite differences on the smooth RBF interpolation of the summary statistic $\bb{d}(\bb{p})$. This linear approximation allows for a fast estimate of $\bbh{p}$ in terms of $\bbh{d}_{\rm obs}$
\begin{equation}
\label{meth:parestimate}
\bbh{p} = \bb{p}_0 + (\bb{M}^T\bbh{\Psi}\bb{M})^{-1}\bb{M}^T\bbh{\Psi}(\bbh{d}_{\rm obs}-\bb{d}(\bb{p}_0))
\end{equation}
Here $\bbh{\Psi}\equiv\bbh{C}^{-1}$ denotes the summary statistic's precision matrix. With the linear approximation (\ref{meth:linapprox}) the parameter likelihood (\ref{meth:paramlikelihood}) is a multivariate Gaussian in $\bb{p}$ and its width $\bbh{\Sigma}$ can be estimated as 
\begin{equation}
\label{meth:parcovestimator}
(\bbh{\Sigma})_{\alpha\beta} = -\left(\frac{\partial^2 \log \mathcal{L}(\bb{p})}{\partial p_\alpha \partial p_\beta}\right)^{-1}_{\bb{p}_0} = ((\bb{M}^T\bbh{\Psi}\bb{M})^{-1})_{\alpha\beta}
\end{equation}
The square of the $1\sigma$ parameter errors $\Delta \bbh{p}^2$ are the diagonal entries of $\bbh{\Sigma}$. The parameter covariance estimator (\ref{meth:parcovestimator}) is the same as one gets adopting a Fisher Matrix formalism for parameter forecasts (see \citep{astroMLText}).

When the dimension $N_b$ of the summary statistics space is large, numerical issues can arise in the estimation of the parameter error bars if the covariance matrix $\bbh{C}$ is measured from simulations. When $N_r$ independent realizations are used to estimate $\bbh{C}$, its inverse $\bbh{\Psi}$ is biased by a constant factor (see \citep{Hartlap07,Taylor12,Taylor14}) which can be taken into account. When the bias correction is applied, we can calculate the expectation value of the covariance estimator (\ref{meth:parcovestimator}) (see again\citep{Taylor14})
\begin{equation}
\label{meth:covdegradation}
\langle\bbh{\Sigma}\rangle = \left(\frac{N_r-N_b+N_p-1}{N_r-N_b-2}\right)\bb{\Sigma}
\end{equation}   
where $\bb{\Sigma}$ is the asymptotic covariance one obtains with an infinite number of realizations and $N_p=3$ is the number of parameters we are estimating. The scatter of the parameter estimates (\ref{meth:parestimate}) on the other hand scales as \citep{Taylor14}
\begin{equation}
\label{meth:estdegradation}
\langle{\rm Cov}(\bbh{p})\rangle = \left(\frac{N_r-2}{N_r-N_b+N_p-2}\right)\bb{\Sigma}
\end{equation}
Although equations (\ref{meth:covdegradation}) and (\ref{meth:estdegradation}) agree in the limit $N_r\rightarrow\infty$, they can be different when a finite number of realizations is used. The degradation factor in the parameter covariance estimate in (\ref{meth:covdegradation}) is of order $1+(1+N_p)/N_r$, while the scatter of the estimates $\bbh{p}$ is of order $1+(N_b-N_p)/N_r$. These numbers can be very different if $N_b$ is large. This means that the parameter error bar estimate (\ref{meth:covdegradation}) is too conservative if $N_b/N_r$ is of order unity. This could be the case with the inclusion of tomography information. If we bin the single redshift summary statistic with $N_{\rm st}$ intervals, and consider $N_z$ redshift bins, this can lead to a summary statistic vector of size $N_b = O(N_{\rm st}N_z^2)$ for the power spectrum and $N_b = O(N_{\rm st}N_z)$ for the remaining statistics. This can become quickly comparable with $N_r=16000$ once more redshift bins or a finer binning of the summary statistic are considered. In order to avoid these error degradation issues, we apply dimensionality reduction techniques to the summary statistics we are considering. Even if these techniques might not play a vital role in this work, as the maximum $N_b/N_r$ ratio we use is of order 1\%, they will definitely be relevant in future experiments when using finely binned summary statistics or when combining different cosmological probes. We explain the dimensionality reduction techniques we adopted in the next paragraph.


\subsection{Dimensionality reduction}
We apply a Principal Component Analysis (see \citep{astroMLText} for example) to reduce the dimensionality of our summary statistics while preserving the cosmological information content. The model statistic $\bb{d}(\bb{p})$ can be regarded as a $P\times N_b$ matrix $d_{pi}$. Consider the whitened model matrix
\begin{equation}
\label{meth:whitening}
\Delta_{pi} = P\frac{d_{pi}}{\sum_{q=1}^Pd_{qi}} - 1.
\end{equation} 
We perform a Singular Value Decomposition (SVD) of $\Delta$ 
\begin{equation}
\bb\Delta = \bb{L}\bb{\Lambda} \bb{R}
\end{equation}
where $L$ is $P\times Q$, $\bb{\Lambda}={\rm diag}(\Lambda_1,...,\Lambda_Q)$ and $\bb{R}$ is $Q\times N_b$ and $Q={\min}(P,N_b)$. $R_{ni}$ is the $i$-th component of the $n$-th basis vector in statistics space. The singular value $\Lambda_n$ is the variance of the whitened summary statistic along the $n$-th basis vector. We assume that only summary statistic projections on the first $N_c$ basis vectors contain relevant cosmological information, where $N_c<N_b$ is a number that has to be determined from the simulations. Let $\bb{R}(N_c)$ be a matrix made of the first $N_c$ rows of $\bb{R}$ (we assume that the singular values $\Lambda_i$ are sorted from highest to lowest). We define the PCA projection of a summary statistic $\bbh{d}$ on $N_c$ principal components as 
\begin{equation}
\label{meth:pcaprojection}
\h{d}_n(N_c) = \sum_{i=1}^{N_b}(\bb{R}(N_c))_{ni}\left(P\frac{\h{d}_i}{\sum_{p=1}^P d_{pi}}-1\right) 
\end{equation} 
Through the above procedure, we hope to capture the cosmological information contained in $\bbh{d}$ by projecting it on the $N_c<N_b$ principal components that vary the most with cosmology parameters.  


\subsection{Priors from CMB experiments}
In this paragraph we describe how we included prior knowledge of cosmological parameters from previous Cosmic Microwave Background (CMB) observations, such as Planck \citep{PlanckCosmo}. CMB experiments provide tight constraints on $(\Omega_m,\sigma_8)$, but they are not sensitive to dark energy parameters such as $w$. Nevertheless, prior knowledge of $\Omega_m$ and $\sigma_8$ could in principle help in breaking degeneracies between these parameters and $w$ in weak lensing observations. The CMB parameter prior probability function can be written as 
\begin{equation}
-2\log\mathcal{L}_{\rm CMB}(\bb{p}) = (\bb{p} - \bb{p}_0)^T\bb{F}_{\rm CMB}(\bb{p} - \bb{p}_0)
\end{equation}
where we assumed that the best fit parameters are the same $\bb{p}_0$ that appear in equation (\ref{meth:linapprox}). 
Parameter constraints from the Planck CMB experiment are made available to the public via the parameter Markov Chains (MCMC) published on the Planck Legacy Archive \footnote{The archive we used is located \url{http://pla.esac.esa.int/pla/}; we used the MCMC chains contained in the \ttt{base\_w/plikHM\_TT\_lowTEB} directory, labeled as \ttt{base\_w\_plikHM\_TT\_lowTEB\_[1-4].txt}}. We can use these MCMC data to estimate the parameter covariance matrix $\bb{\Sigma}_{\rm CMB}$ on the parameter multiplet $(\Omega_m,\Omega_bh^2,\theta_{\rm MC},\tau,w,n_s,\sigma_8)$, marginalized over the Planck nuisance parameters. We then compute the parameter prior Fisher matrix $\bb{F}_{\rm CMB} = \bb{\Sigma}_{\rm CMB}^{-1}$. Fixing the values of all parameters but $(\Omega_m,w,\sigma_8)$ and applying the prior to the weak lensing parameter likelihood (\ref{meth:paramlikelihood}) is equivalent to taking the $(\Omega_m,w,\sigma_8)$ slice of $\bb{F}_{\rm CMB}$, which we call $\bb{F}_{\rm CMB}^{(\Omega_m,w,\sigma_8)}$, and computing the parameters constraints subject to the CMB prior as 

\begin{equation}
\label{meth:parcovestimator_cmb}
\bbh{\Sigma}_{\rm lens+CMB} = \left(\bb{M}^T\bbh{\Psi}\bb{M} + \bb{F}_{\rm CMB}^{(\Omega_m,w,\sigma_8)}\right)^{-1}
\end{equation}
In the next section we describe the main results of this work. 


\section{Results}
\label{sec:results}

\begin{figure*}
\includegraphics[scale=0.3]{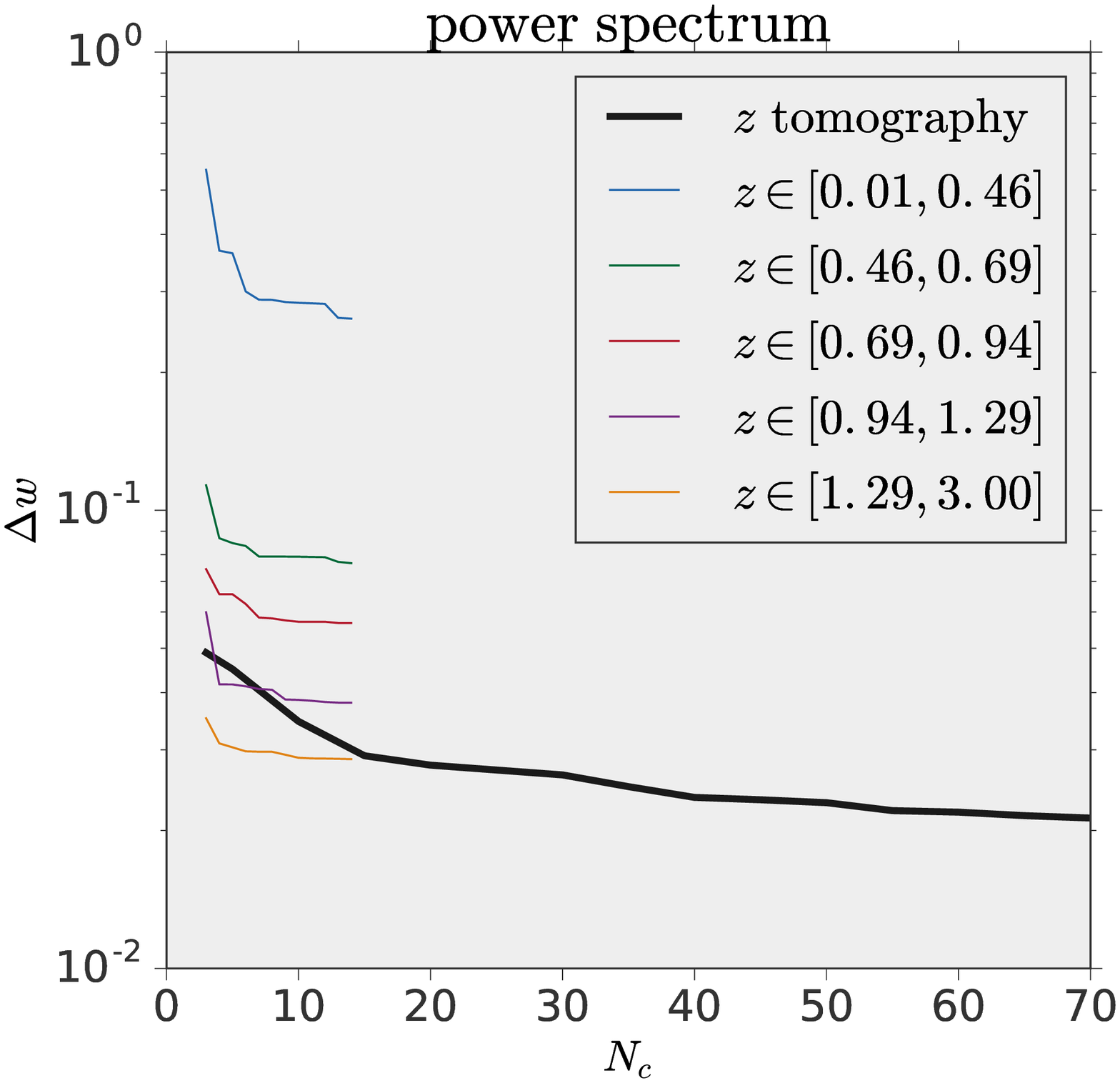}
\includegraphics[scale=0.3]{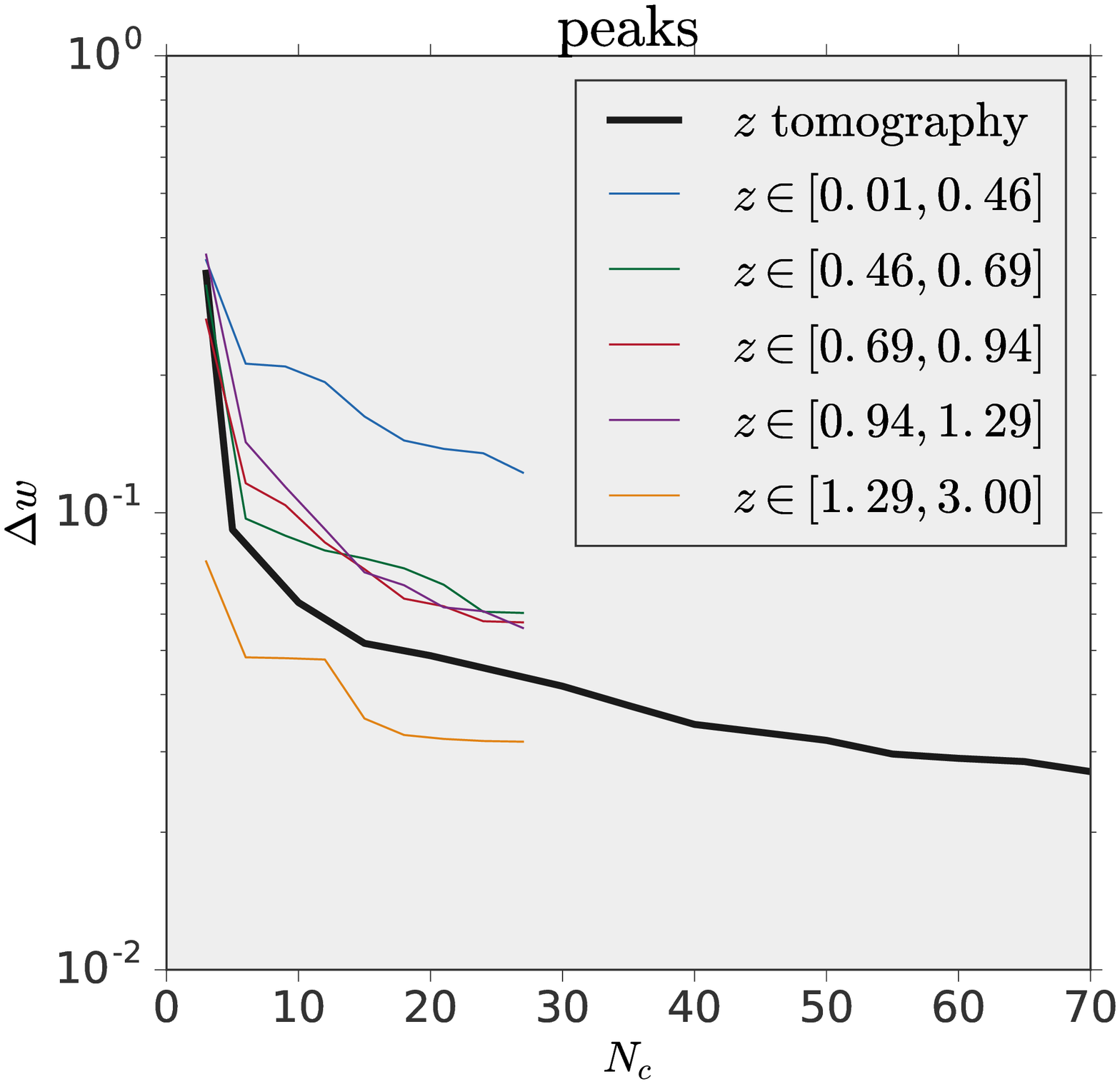}
\includegraphics[scale=0.3]{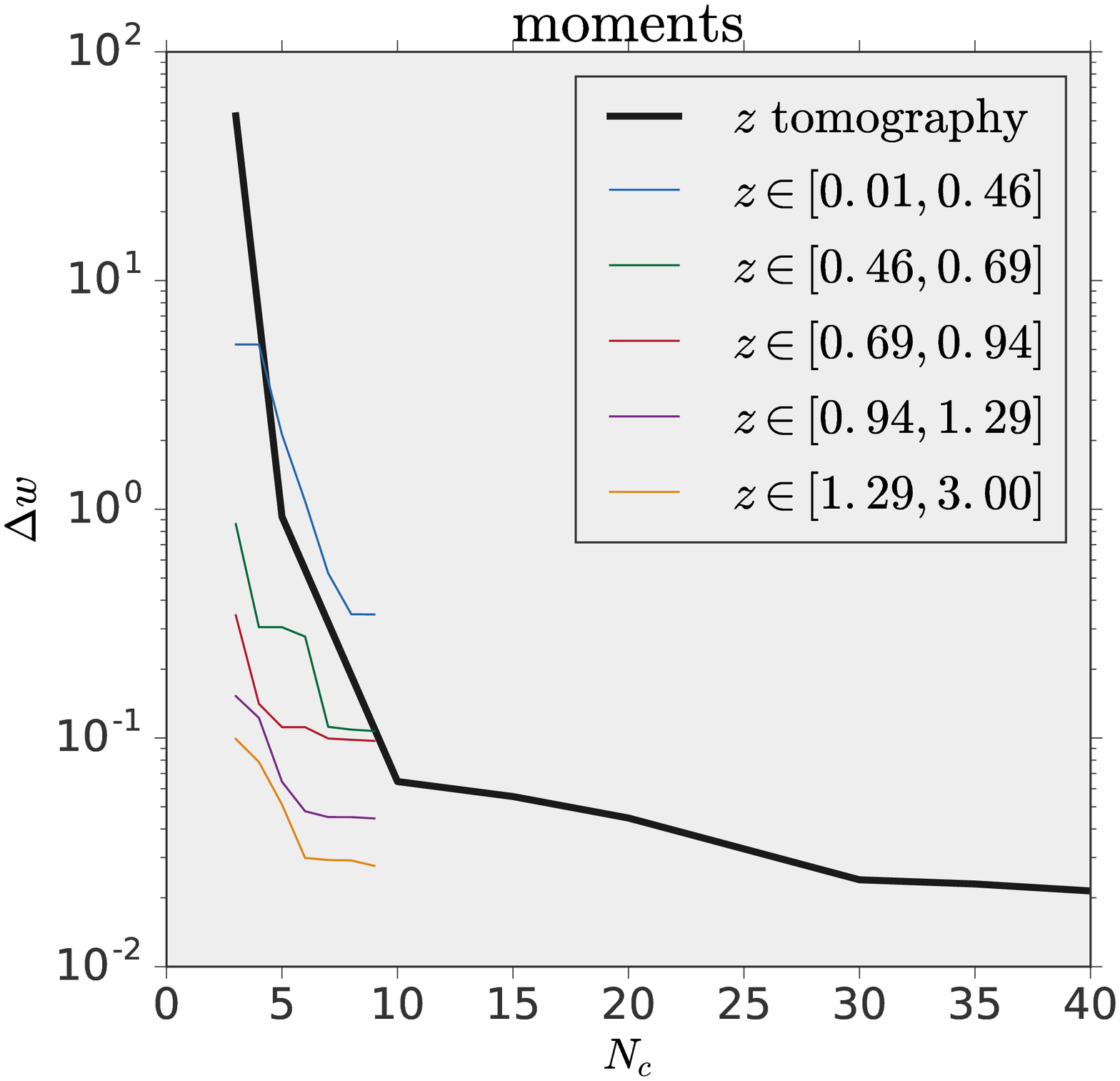}
\caption{$1\sigma$ errors on $w$, marginalized over $(\Omega_m,\sigma_8)$, as a function of the number of the principal components $N_c$, using the power spectrum (top left), peak counts (top right) and moments (bottom). The thin colored lines refer to single redshift summary statistics, while the thick black line shows the case in which the joint redshift information is included.}
\label{fig:pcacomponents}
\end{figure*}

In this section we present the main results of this work. Figure \ref{fig:pcacomponents} shows the robustness of the dimensionality reduction technique we adopted for the three summary statistics considered, namely the convergence power spectrum $P^{\kappa\kappa}(\ell,\bar{z}_b,\bar{z}_{b'})$, peak counts $n_{\rm pk}(\nu,\bar{z}_b)$ and moments $\pmb{\mu}(\bar{z}_b)$. To measure the power spectrum we chose 15 uniformly spaced multipole bands between $(\ell_{\rm min},\ell_{\rm max})=(200,2000)$. There are only 15 independent $(\bar{z}_b,\bar{z}_{b'})$ combinations (5 diagonal + 10 off-diagonal), which leads to a total of $N_b=15({\rm multipoles})\times 15({\rm redshift})=225$ power spectrum measurement bands, including cross redshift information. 
We bin the convergence peak counts in 30 uniformly spaced $\nu$ bins between $(\nu_{\rm min},\nu_{\rm max})=(-2,7)$, for a total of $N_b=30 (\rm peak \,\,heights)\times5({\rm redshift})=150$ measurement bands. The total size of the moments summary statistic vector is $N_b=9(\rm moments)\times 5(\rm redshift)=45$. 

The forecast error bars on $w$ are calculated according to equation (\ref{meth:parcovestimator}), where the covariance matrix $\bbh{C}$ and its inverse $\bbh{\Psi}$ have been estimated from $N_r=16000$ realizations of each summary statistic in the fiducial cosmology.

\begin{figure*}
\includegraphics[scale=0.3]{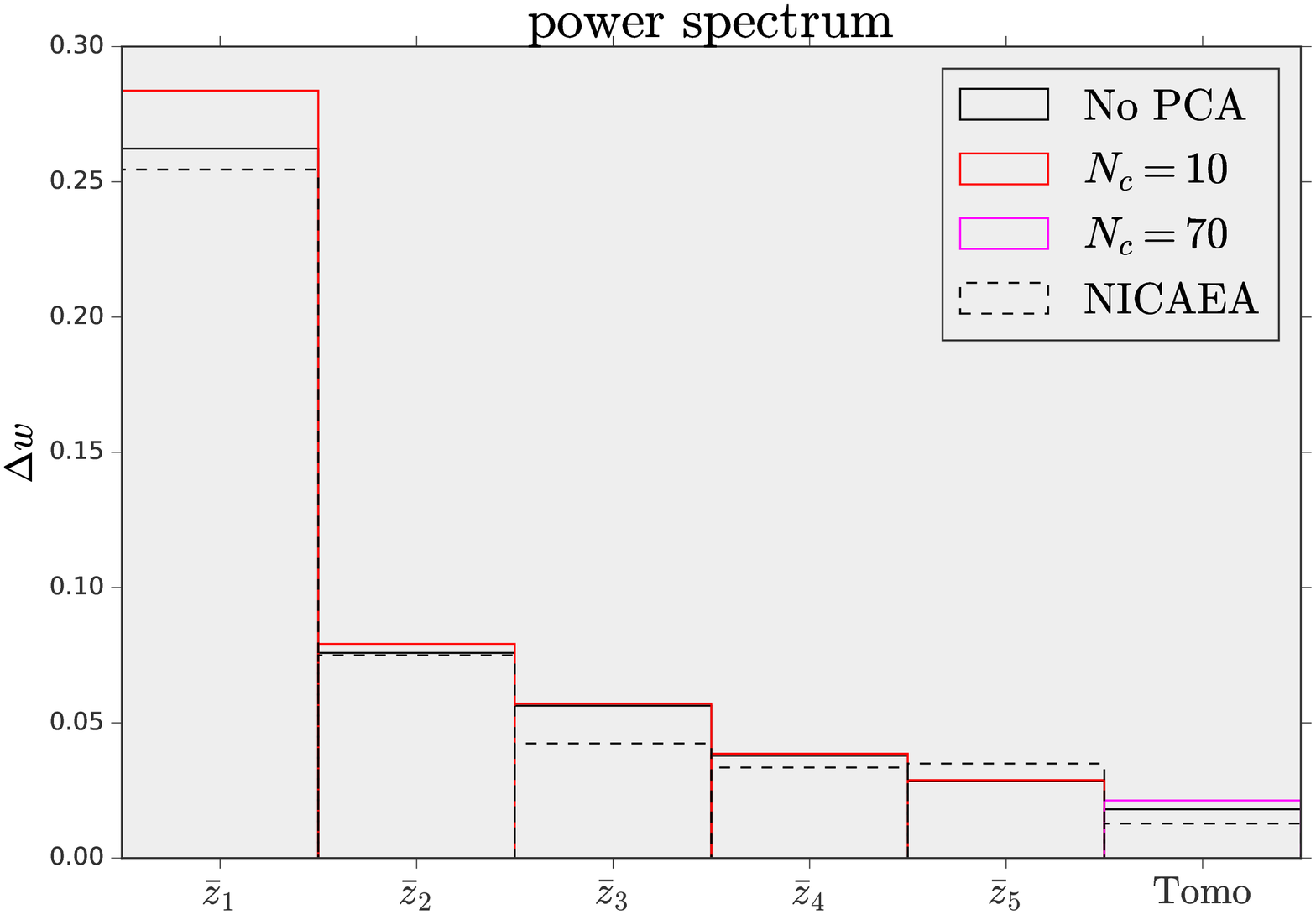}
\includegraphics[scale=0.3]{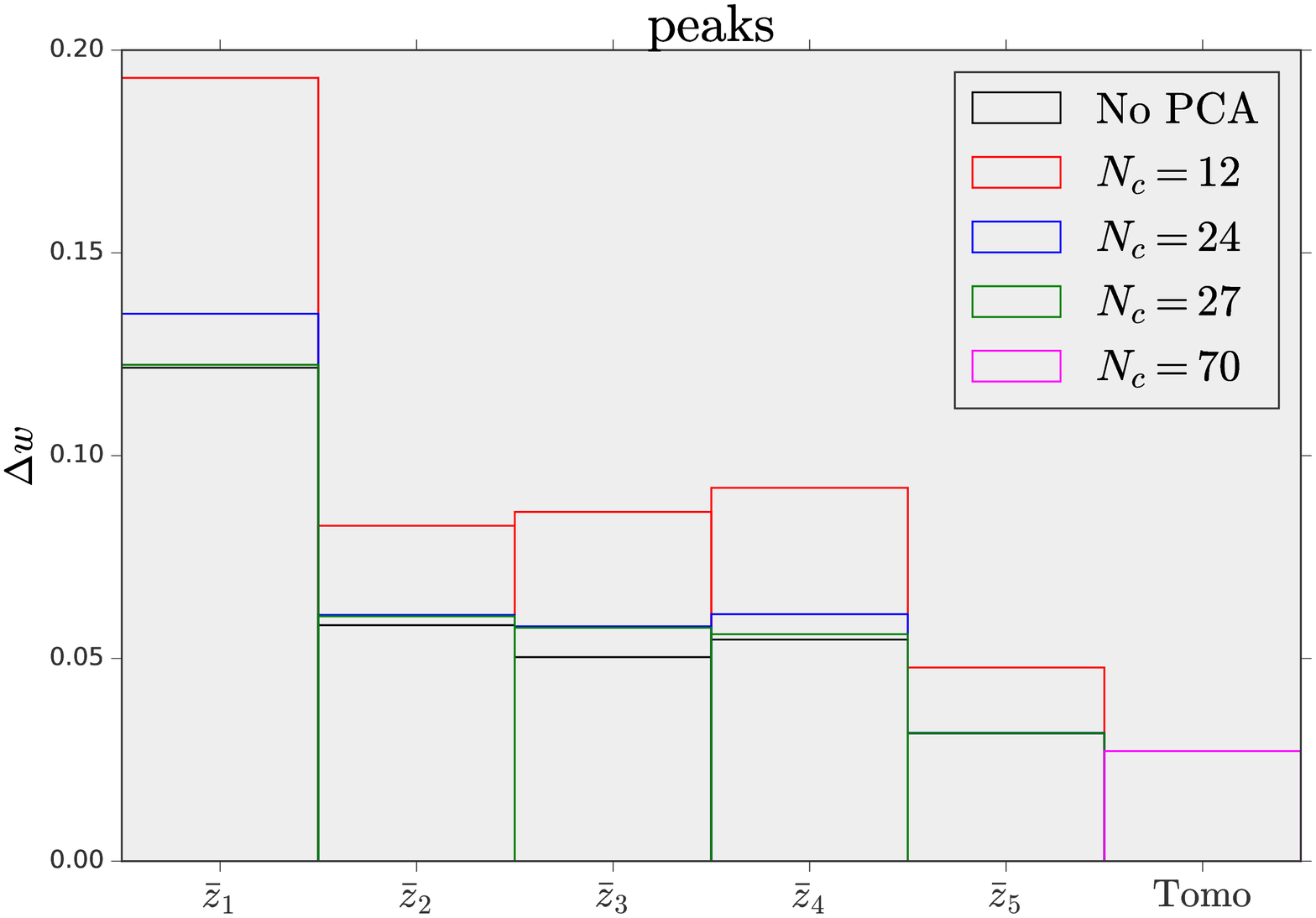}
\includegraphics[scale=0.3]{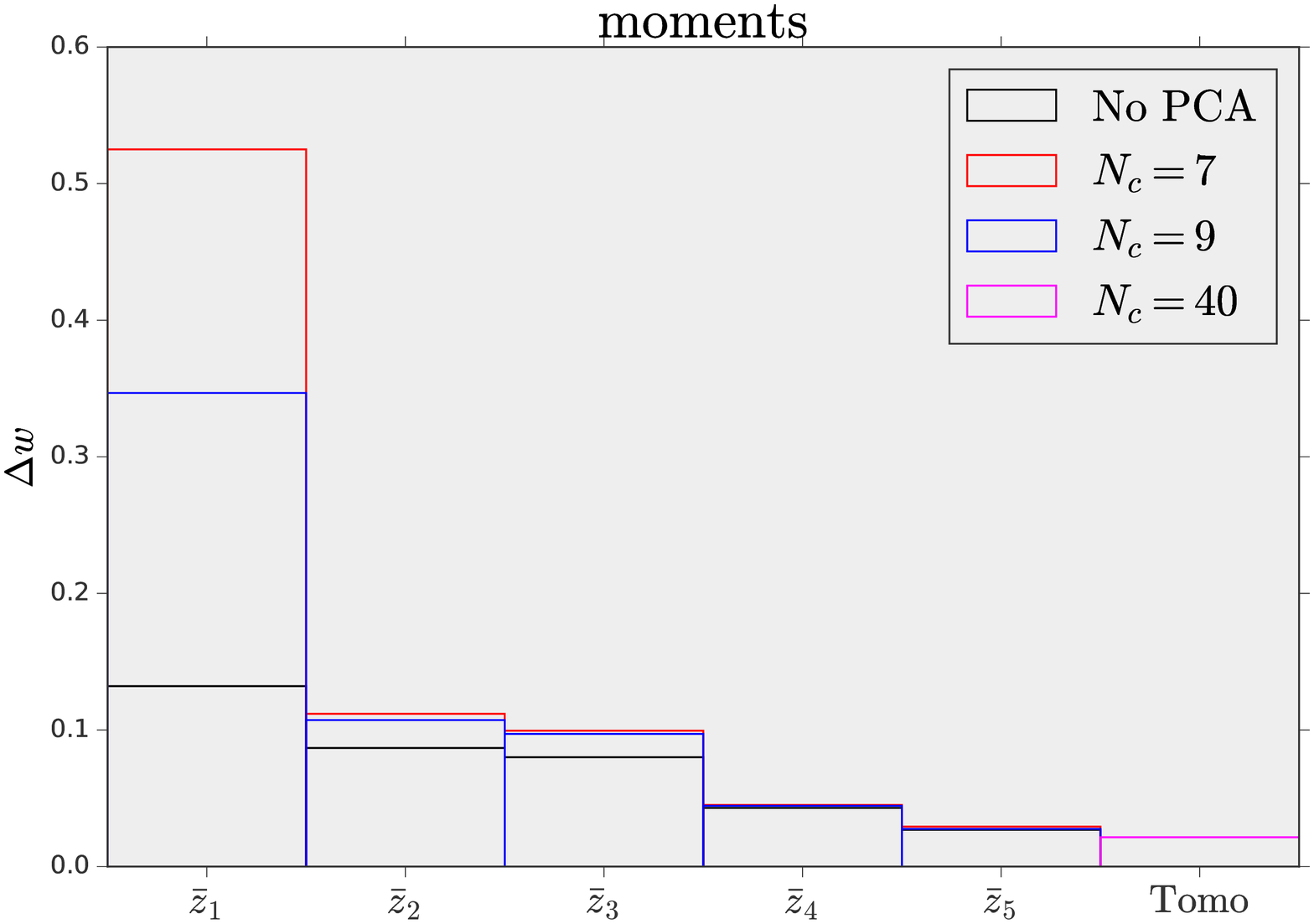}
\caption{Comparison between single-redshift $w$ constraints (marginalized over $\Omega_m$ and $\sigma_8$) obtained without PCA (black bars) and with PCA (colored bars) as a function of the redshift bin $\bar{z}_b$. We show constraints obtained from the power spectrum (top left), peak counts (top right) and moments (bottom). Note the different scales on the vertical axis in the three different panels. The black dashed lines in the top left panel refer to parameter constraints on $w$ obtained with the public code NICAEA \citep{Nicaea}, assuming a Gaussian covariance model for the power spectrum, as specified by equation (\ref{meth:gausspcov}).}
\label{fig:nopca}
\end{figure*}

Figure \ref{fig:nopca} shows a comparison between the $w$ constraints obtained using single redshift bins, with and without PCA dimensionality reduction, and compares these single redshift constraints with the ones obtained using redshift tomography. When we calculate parameter inferences using the convergence power spectrum $P^{\kappa\kappa}$, we can cross check the results obtained with our simulations with the ones obtained with the analytical code NICAEA \citep{Nicaea}. This code allows to predict the convergence power spectrum as a function of cosmological parameters $\bb{p}$, for an arbitrary galaxy redshift distribution $n(z)$. Parameter inferences can be obtained from the NICAEA predictions for $P^{\kappa\kappa}(\ell,z_i,z_j)$ (where $\{z_i\}$ are the centers of the redshift bins) using equation (\ref{meth:parcovestimator}). To proceed in the calculations we approximate the $P^{\kappa\kappa}$ covariance matrix with the one one would obtain in the limit in which the $\kappa(\pmb{\theta})$ field is Gaussian

\begin{widetext}
\begin{equation}
\label{meth:gausspcov}
\left\langle\delta\h{P}_\ell(z_1,z_2)\delta\h{P}_\ell(z_3,z_4)\right\rangle = \frac{P_\ell(z_1,z_3)P_\ell(z_2,z_4)+P_\ell(z_1,z_4)P_\ell(z_2,z_3)}{f_{\rm sky}\delta\ell_{\rm bin}(2\ell+1)}
\end{equation} 
\end{widetext}
where $P_\ell(z_1,z_2)$ is a shorthand for $P^{\kappa\kappa}(\ell,z_1,z_2)$, $\delta\h{P}_\ell(z_1,z_2)=\h{P}_\ell(z_1,z_2) - P_\ell(z_1,z_2)$ is the scatter in the $\h{P}$ estimator, $\delta\ell_{\rm bin}$ is the width of the linearly spaced multipole bands and $f_{\rm sky} = \theta^2_{\rm FOV}/4\pi$ is the sky coverage fraction of one field of view. In this approximation the cross variance terms between different multipoles are assumed to be zero.   

\begin{figure*}
\includegraphics[scale=0.3]{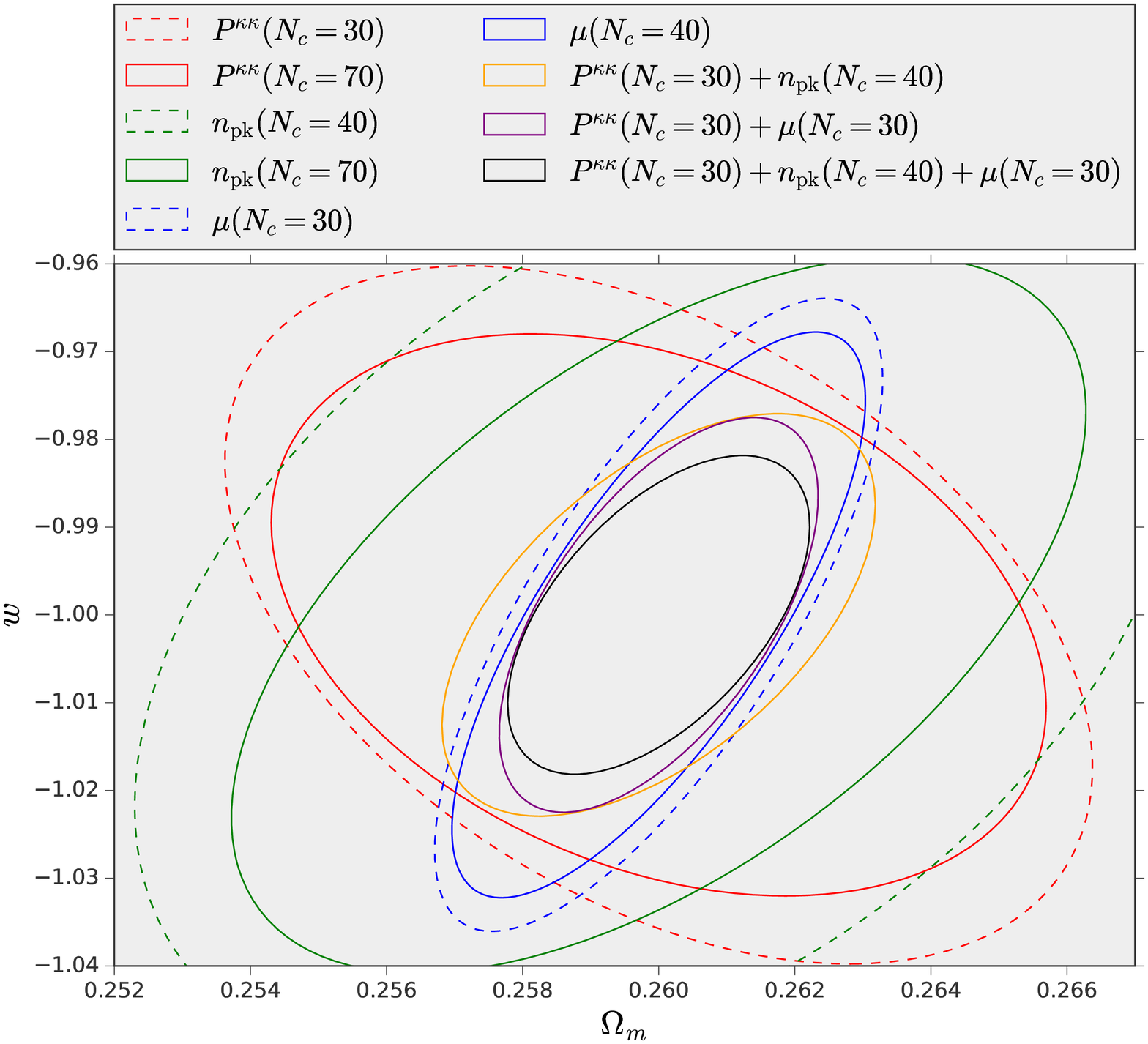}
\includegraphics[scale=0.3]{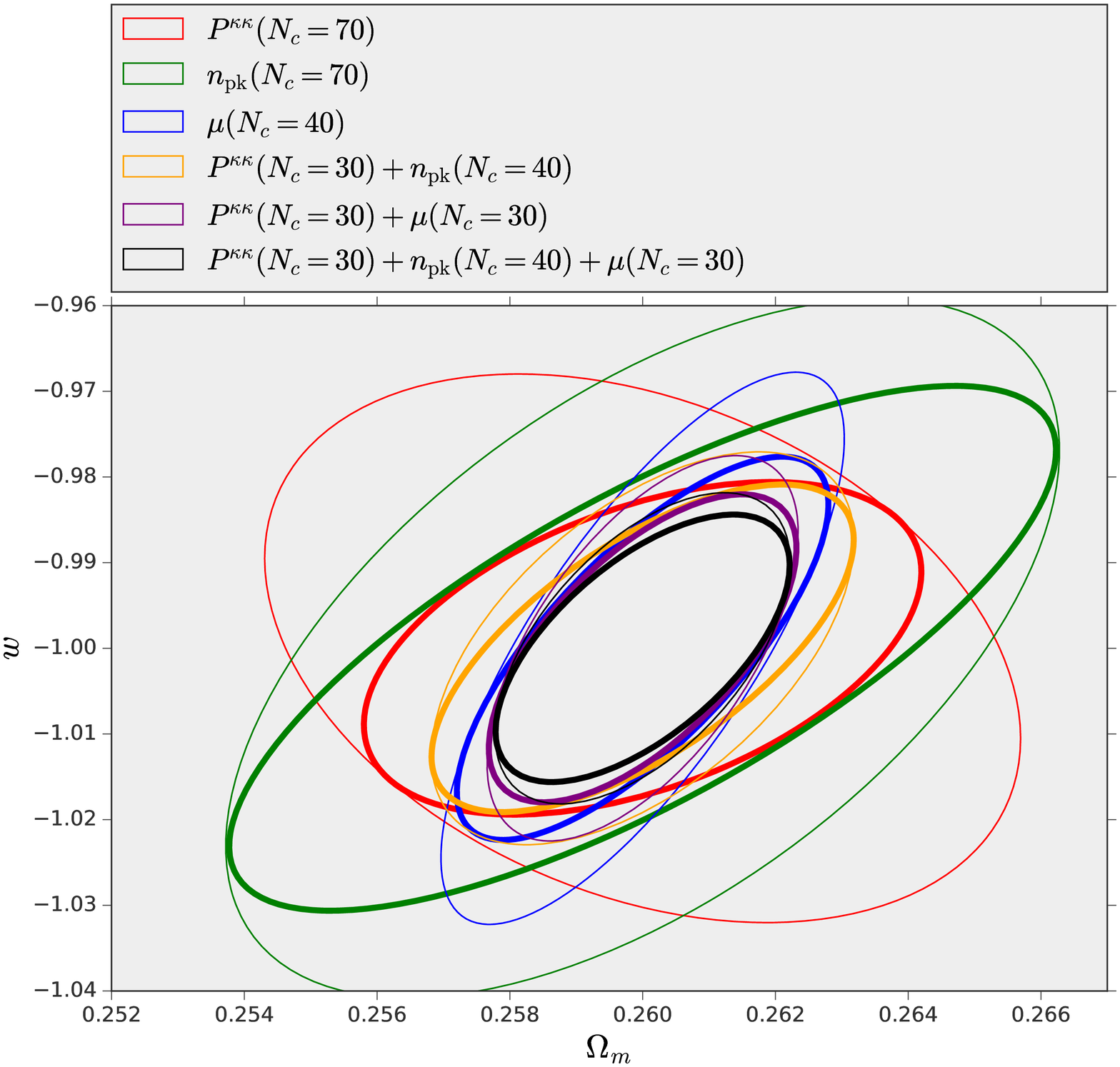}
\caption{$1\sigma$ tomographic constraints on the $(\Omega_m,w)$ parameter space, marginalized over $\sigma_8$, obtained using equation (\ref{meth:parcovestimator}). The covariance matrix $\bbh{C}$ and its inverse $\bbh{\Psi}$ have been computed from $16000$ summary statistics realizations, and have been scaled by a factor $N_{\rm FOV}=1600$ to mimic the constraining power of an LSST-sized survey. The thick line ellipses in the right panel refer to $(\Omega_m,w)$ obtained from the weak lensing statistics considered in this work, but subject to Planck priors as described in equation (\ref{meth:parcovestimator_cmb}). The thin solid lines in the left and right panels are the same.}
\label{fig:constraintsOm-w}
\end{figure*}

Figure \ref{fig:constraintsOm-w} shows the $1\sigma$ confidence contours on the $(\Omega_m,w)$ doublet calculated from equation (\ref{meth:parcovestimator}) after the PCA dimensionality reduction performed according to equation (\ref{meth:pcaprojection}), for a variety of choices of statistic and $N_c$. We also show the improvements on the confidence contours when combining different summary statistics after the corresponding dimensionality reductions have been performed. The constraints in the $(\Omega_m,w,\sigma_8)$ parameter space for a variety of summary statistics are summarized in Tables \ref{tbl:constraints} (weak lensing only) and \ref{tbl:constraints-cmb} (with priors from Planck added). 

\begin{figure}
\includegraphics[scale=0.3]{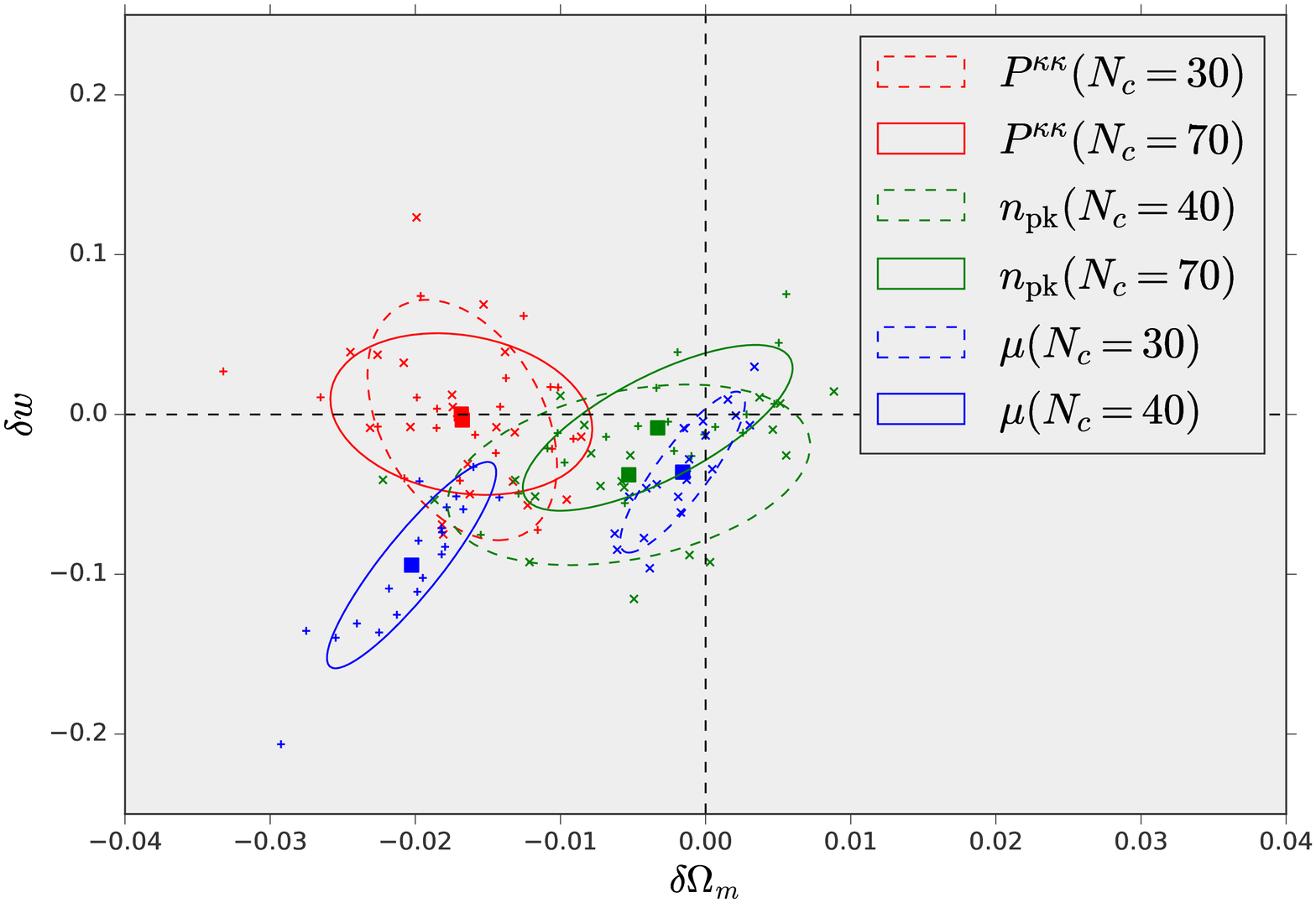}
\caption{Distribution of parameter estimates in the $(\Omega_m,w)$ parameter space using the power spectrum (red), peak counts (green) and moments (blue) to fit 20 LSST-like simulated observations. The parameter deviations $(\delta\Omega_m,\delta w)$ are obtained comparing parameter estimates from equation (\ref{meth:parestimate}) with and without photo-$z$ errors. The colored squares and the ellipses correspond to respectively to the mean and $1\sigma$ level of the $(\delta\Omega_m,\delta w)$ distribution, assuming the latter is Gaussian.}
\label{fig:photozbias}
\end{figure}

Figure \ref{fig:photozbias} shows the effect of ignoring photo-$z$ errors on parameter constraints. To evaluate this effect we construct different simulated observations, with and without photo-$z$ errors, and compare the results of the parameter fit according to equation (\ref{meth:parestimate}). Using our simulation suite, we construct 20 simulated observations: the summary statistic in each observation is calculated by taking the mean of a random sample of $N_{\rm FOV}=1600$ realizations of the summary statistic in the fiducial cosmology (randomly chosen among the ensemble of $N_r=16000$ that are available in the ensemble). The estimated covariance matrix $\bbh{C}$ is scaled by a factor $N_{\rm FOV}$ to take into account the construction process of the simulated observations. This procedure allows us to forecast the results an LSST-like survey would obtain. We stress that, because of the small size of our simulation box the covariance estimate that we obtain is likely not accurate enough to produce constraints from LSST data. Full treatment of observables covariance matrices, along with larger $N$--body simulations and SSC effects will be investigated in future work.      

\begin{table*}
\begin{tabular}{c|c|c|c|c|c|}
\toprule
                                      \textbf{Statistic} & $\Delta \Omega_m$ & $\Delta w$ & $\Delta \sigma_8$ & $10^6{\rm Area} (\Omega_m,w)$ & $10^9{\rm Volume}$ $(\Omega_m,w,\sigma_8)$ \\ \hline \hline
\midrule
                   Power spectrum ($\bar{z}_5$) &            0.0222 &     0.0286 &            0.0298 &                           632 &                                        654 \\
                          Power spectrum (tomo) &            0.0038 &     0.0213 &            0.0060 &                            76 &                                         74 \\
                            Peaks ($\bar{z}_5$) &            0.0049 &     0.0316 &            0.0050 &                            98 &                                         99 \\
                                   Peaks (tomo) &            0.0042 &     0.0271 &            0.0043 &                            93 &                                        122 \\
                          Moments ($\bar{z}_5$) &            0.0027 &     0.0276 &            0.0026 &                            48 &                                         39 \\
                                 Moments (tomo) &            0.0020 &     0.0214 &            0.0020 &                            28 &                                         21 \\ \hline \hline
           
           Power spectrum + peaks ($\bar{z}_5$) &            0.0040 &     0.0209 &            0.0044 &                            58 &                                         53 \\
                  Power spectrum + peaks (tomo) &            0.0021 &     0.0153 &            0.0026 &                            27 &                                         26 \\
         Power spectrum + moments ($\bar{z}_5$) &            0.0023 &     0.0190 &            0.0025 &                            32 &                                         26 \\
                Power spectrum + moments (tomo) &            0.0016 &     0.0150 &            0.0019 &                            18 &                                         14 \\
 Power spectrum + peaks + moments ($\bar{z}_5$) &            0.0020 &     0.0127 &            0.0024 &                            21 &                                         17 \\
        Power spectrum + peaks + moments (tomo) &            0.0015 &     0.0121 &            0.0018 &                            14 &                                         11 \\ \hline
\bottomrule
\end{tabular}
\caption{Constraints on the $(\Omega_m,w,\sigma_8)$ parameter triplet using different summary statistics and redshift information. Each column $(\Delta \Omega_m,\Delta w,\Delta \sigma_8)$ contains the 1$\sigma$ constraints on a particular parameter, marginalized over the other two. The last two columns contain respectively the area of the $(\Omega_m,w)$ 68\% confidence level ellipse (marginalized over $\sigma_8$) and the volume of the 68\% confidence level ellipsoid in $(\Omega_m,w,\sigma_8)$ space, both calculated as the square root of the determinant of the relevant $\bbh{\Sigma}$ minors.}
\label{tbl:constraints}
\end{table*}

\begin{table*}
\begin{tabular}{c|c|c|c|c|c|}
\toprule
                                      \textbf{Statistic} & $\Delta \Omega_m$ & $\Delta w$ & $\Delta \sigma_8$ & $10^6{\rm Area} (\Omega_m,w)$ & $10^9{\rm Volume}$ $(\Omega_m,w,\sigma_8)$ \\ \hline \hline
\midrule
                   Power spectrum ($\bar{z}_5$) &            0.0076 &     0.0274 &            0.0084 &                            94 &                                         96 \\
                          Power spectrum (tomo) &            0.0028 &     0.0129 &            0.0035 &                            32 &                                         31 \\
                            Peaks ($\bar{z}_5$) &            0.0048 &     0.0237 &            0.0050 &                            57 &                                         57 \\
                                   Peaks (tomo) &            0.0041 &     0.0204 &            0.0042 &                            55 &                                         70 \\
                          Moments ($\bar{z}_5$) &            0.0024 &     0.0172 &            0.0025 &                            27 &                                         21 \\
                                 Moments (tomo) &            0.0019 &     0.0149 &            0.0020 &                            18 &                                         14 \\ \hline \hline

           Power spectrum + peaks ($\bar{z}_5$) &            0.0040 &     0.0184 &            0.0043 &                            40 &                                         36 \\
                  Power spectrum + peaks (tomo) &            0.0021 &     0.0127 &            0.0025 &                            20 &                                         19 \\
         Power spectrum + moments ($\bar{z}_5$) &            0.0022 &     0.0145 &            0.0025 &                            22 &                                         17 \\
                Power spectrum + moments (tomo) &            0.0015 &     0.0120 &            0.0018 &                            14 &                                         11 \\
 Power spectrum + peaks + moments ($\bar{z}_5$) &            0.0019 &     0.0110 &            0.0023 &                            16 &                                         13 \\
        Power spectrum + peaks + moments (tomo) &            0.0015 &     0.0104 &            0.0018 &                            12 &                                          9 \\ \hline
\bottomrule
\end{tabular}
\caption{Constraints on the $(\Omega_m,w,\sigma_8)$ parameter triplet using different summary statistics and redshift information, including Fisher priors from Planck according to equation (\ref{meth:parcovestimator_cmb}). Each column $(\Delta \Omega_m,\Delta w,\Delta \sigma_8)$ contains the 1$\sigma$ constraints on a particular parameter, marginalized over the other two. The last two columns contain respectively the area of the $(\Omega_m,w)$ 68\% confidence level ellipse (marginalized over $\sigma_8$) and the volume of the 68\% confidence level ellipsoid in $(\Omega_m,w,\sigma_8)$ space, both calculated as the square root of the determinant of the relevant $\bbh{\Sigma}$ minors.}
\label{tbl:constraints-cmb}
\end{table*}

\section{Discussion}
\label{sec:discussion}

In this section we discuss our findings. Figure \ref{fig:pcacomponents} shows that our dimensionality reduction technique is robust. In particular, for all the summary statistics we consider, a plateau in the $w$ error is reached for a sufficient high number of principal components $N_c$. We also see that for single redshift statistics, this plateau is reached for $\sim 5$ components for the power spectrum and the moments, and $\sim 10$ components for the peak counts. Moreover, Figure \ref{fig:nopca} shows that, at least for the four highest redshift bins $\{\bar{z}_b: b\geq 2\}$, most of the cosmological information contained in the full (pre-PCA) summary statistic vector can be captured with a limited number of principal components $N_c<N_b$. The minimum number of components necessary to capture most of the available information increases when including redshift tomography, and can reach $\sim 30$ for the power spectrum and moments and $\sim 40$ for the peak counts. 

Figure \ref{fig:nopca} also clearly shows that, when considering a single redshift bin and a single summary statistic, most of the information on $w$ is contained in the highest redshift galaxies. PCA does not seem to capture all the information in the lowest redshift bin, even when enough components are included to reach the plateau in Figure \ref{fig:pcacomponents}. This can be attributed to the fact that PCA is not scale-invariant \citep{astroMLText}, because there is freedom in choosing the whitening scale in equation (\ref{meth:whitening}). Our choice of the whitening scale seems to affect significantly the first redshift bin, with the effect being mitigated for the highest redshift bins. 
The top left panel of Figure \ref{fig:nopca} also shows reasonable agreement between the results we obtain with our simulations and the ones we calculate with the analytical code NICAEA. 

There are two possibilities for improving the constraints: the use of redshift tomography and the combination of different statistics. Table \ref{tbl:constraints} shows that the area and volume of the $(\Omega_m,w)$ ellipse and $(\Omega_m,w,\sigma_8)$ ellipsoid shrink by a factor of 8 when redshift tomography is added to the power spectrum, while the improvement is more modest for the remaining statistics (negligible for the peaks, and a factor of 2 for the moments). Combining the power spectrum and the peak counts in the highest redshift bin leads to a factor of 10 improvement in the $(\Omega_m,w)$ and $(\Omega_m,w,\sigma_8)$ 68\% confidence intervals, with tomography further shrinking the contours by an additional factor of 2. Combining the power spectrum and the moments in the highest redshift bin provides 20 times tighter constraints on $(\Omega_m,w)$ and $(\Omega_m,w,\sigma_8)$, with tomography yielding an additional factor of 2 improvement. Table \ref{tbl:constraints} also shows that power spectrum tomography can help breaking the degeneracy between $\Omega_m$ and $\sigma_8$. The same is not true for peaks and moments tomography, although combining these statistics with the power spectrum yields a factor of respectively 2 and 3 better constraints on $\Omega_m$ and $\sigma_8$. 

Table \ref{tbl:constraints-cmb} shows that parameter priors from Planck yield a factor of 6 improvement on the $(\Omega_m,w)$ and $(\Omega_m,w,\sigma_8)$ 68\% confidence intervals, even when a single redshift bin is considered. When the Planck priors in equation (\ref{meth:parcovestimator_cmb}) are included, the improvements in constraints when adding redshift tomography or combining different statistics are more moderate. Tomography improves power spectrum constraints by a factor of 3. Adding moments improves by an additional factor of 2, and  adding both moments and peaks improves by almost a factor of 3 over power spectrum tomography alone. 

Figure \ref{fig:constraintsOm-w} shows that peaks and moments contain cosmological information that is not contained in the power spectrum, because a similar improvement cannot be obtained by simply increasing the number of PCA components in the power spectrum dimensionality reduction procedure. This is consistent with previous work (see for example \citep{CFHTMink})

Figure \ref{fig:photozbias} quantifies the effect of uncorrected photo-$z$ errors on the $(\Omega_m,w)$ constraints. Because the stochastic nature of the observations, parameter values $\bbh{p}$ estimated from equation (\ref{meth:parestimate}) are affected by statistical fluctuations. In Figure \ref{fig:photozbias} we show 20 random draws from the probability distribution of $\delta \bbh{p}=\bbh{p}_{\rm photo-z}-\bbh{p}_{\rm no-photo-z}$. We can conclude that the $\bbh{p}$ estimator is biased if $\langle\delta\bbh{p}\rangle\neq 0$. Figure \ref{fig:photozbias} clearly show that photo-$z$ errors cause parameter biases at more than $1\sigma$ significance level when using the power spectrum and the moments, while no bias is observed for the peak count statistic within its 68\% confidence region. Peak histogram shapes are more robust to this kind of systematic effect since the peak locations are determined by the information coming from neighboring galaxies, while the photo-z errors have no spatial correlation. Photo-z errors are more likely to alter point estimates of the $\kappa$ distribution and larger scale correlations which affect the power spectrum.

We also observe that photo-$z$ errors bias the constraints in slightly different directions, leaving open the possibility of identifying and correcting this bias through self-calibration techniques.         


\section{Conclusions}
\label{sec:conclusions}

In this work we have studied cosmological parameter constraint forecasts for an LSST-like galaxy survey using the convergence power spectrum and a range of non-Gaussian statistics. We make use of redshift tomography to improve the constraints relative to their single-redshift counterparts. We also investigate the effects of uncorrected photo-$z$ systematic effects on the inferred cosmology. Our main findings can be summarized as follows:
\begin{itemize}
	
	\item Principal Component analysis is a robust technique to keep the dimensionality of the parameter space under control and to avoid the numerical pitfalls explained in \citep{Taylor12,DodelsonSchneider13,Taylor14} and more recently in \citep{PetriVariance}. In particular, we find that only a few components (5-10) are necessary to characterize the cosmological information content in single redshift statistics, while more components (30-40) are necessary when tomography is included. Nevertheless we find that the number of required components $N_c$ is significantly smaller than the full summary statistic space dimensionality before performing PCA.
	
	\item When considering a single redshift bin, most of the cosmological information on $w$ is contained in high redshift galaxies. Constraints can be improved with redshift tomography or combining different non-Gaussian statistics with the power spectrum. The improvement originating from the combination of different statistics is attributed to the complementary information that non-Gaussian statistics carry, as a similar improvement cannot be obtained from a single statistic.    
	
	\item Redshift tomography on the power spectrum shrinks the $(\Omega_m,w)$ 68\% confidence ellipse by a factor of 8; combining the peak counts with the power spectrum in the highest redshift bin leads to a factor of 10 better constraints, while adding the moments instead reduces the size of the $(\Omega_m,w)$ ellipse by a factor of 20. When redshift tomography is added on top of these statistics combinations, an additional factor of 2 improvement is observed. Constraint improvements adding redshift tomography and combinations of different statistics are less dramatic when priors from CMB experiments are included in the analysis. 

	\item Uncorrected photo-$z$ systematics can bias parameter constraints obtained from the power spectrum and the moments, but in slightly different parameter directions, leaving open possibilities of somewhat eliminating this bias via self-calibration. 

\end{itemize}

This work explores the advantage of deep galaxy surveys such as LSST, which have access to shape and redshift information of high $z$ galaxies and provide valuable cosmological information on the dark energy equation of state. We also stress the fact that redshift tomography can in some cases provide more stringent constraints on parameters but, for this technique to be viable, accurate knowledge of galaxy redshifts is necessary. Future work needs to address the requirements for photometric measurements accuracy when using non-Gaussian statistics, as well as the self calibration techniques that can be used when different summary statistics are available in addition to the power spectrum. 


\section*{Acknowledgments}

We thank Hu Zhan, Salman Habib, Jeffrey Newman, Colin Hill and Licia Verde for useful discussions. We also thank Martin Kilbinger and Lloyd Knox for comments on an earlier version of this manuscript. Most of the calculations were performed at National Energy Research Scientific Computing Center (NERSC). We thank the LSST Dark Energy Science Collaboration (DESC) for the allocation of time, and for many useful discussions. 
Part of the simulations in this work were also performed at the NSF XSEDE facility, supported by grant number
ACI-1053575, and at the New York Center for Computational Sciences, a
cooperative effort between Brookhaven National Laboratory and Stony
Brook University, supported in part by the State of New York. This
work was supported in part by the U.S. Department of Energy under
Contract No. DE-SC00112704, and by the NSF Grant
No. AST-1210877 (to Z.H.) and by the Research Opportunities and
Approaches to Data Science (ROADS) program at the Institute for Data
Sciences and Engineering at Columbia University (to Z.H.).


\bibliography{ref}

\begin{thebibliography}{47}%
\makeatletter
\providecommand \@ifxundefined [1]{%
 \@ifx{#1\undefined}
}%
\providecommand \@ifnum [1]{%
 \ifnum #1\expandafter \@firstoftwo
 \else \expandafter \@secondoftwo
 \fi
}%
\providecommand \@ifx [1]{%
 \ifx #1\expandafter \@firstoftwo
 \else \expandafter \@secondoftwo
 \fi
}%
\providecommand \natexlab [1]{#1}%
\providecommand \enquote  [1]{``#1''}%
\providecommand \bibnamefont  [1]{#1}%
\providecommand \bibfnamefont [1]{#1}%
\providecommand \citenamefont [1]{#1}%
\providecommand \href@noop [0]{\@secondoftwo}%
\providecommand \href [0]{\begingroup \@sanitize@url \@href}%
\providecommand \@href[1]{\@@startlink{#1}\@@href}%
\providecommand \@@href[1]{\endgroup#1\@@endlink}%
\providecommand \@sanitize@url [0]{\catcode `\\12\catcode `\$12\catcode
  `\&12\catcode `\#12\catcode `\^12\catcode `\_12\catcode `\%12\relax}%
\providecommand \@@startlink[1]{}%
\providecommand \@@endlink[0]{}%
\providecommand \url  [0]{\begingroup\@sanitize@url \@url }%
\providecommand \@url [1]{\endgroup\@href {#1}{\urlprefix }}%
\providecommand \urlprefix  [0]{URL }%
\providecommand \Eprint [0]{\href }%
\providecommand \doibase [0]{http://dx.doi.org/}%
\providecommand \selectlanguage [0]{\@gobble}%
\providecommand \bibinfo  [0]{\@secondoftwo}%
\providecommand \bibfield  [0]{\@secondoftwo}%
\providecommand \translation [1]{[#1]}%
\providecommand \BibitemOpen [0]{}%
\providecommand \bibitemStop [0]{}%
\providecommand \bibitemNoStop [0]{.\EOS\space}%
\providecommand \EOS [0]{\spacefactor3000\relax}%
\providecommand \BibitemShut  [1]{\csname bibitem#1\endcsname}%
\let\auto@bib@innerbib\@empty
\bibitem [{\citenamefont {{Schneider}}(2005)}]{wlreview}%
  \BibitemOpen
  \bibfield  {author} {\bibinfo {author} {\bibfnamefont {P.}~\bibnamefont
  {{Schneider}}},\ }\href@noop {} {\bibfield  {journal} {\bibinfo  {journal}
  {ArXiv Astrophysics e-prints}\ } (\bibinfo {year} {2005})},\ \Eprint
  {http://arxiv.org/abs/astro-ph/0509252} {astro-ph/0509252} \BibitemShut
  {NoStop}%
\bibitem [{\citenamefont {{Heymans}}\ \emph {et~al.}(2012)\citenamefont
  {{Heymans}}, \citenamefont {{Van Waerbeke}}, \citenamefont {{Miller}},
  \citenamefont {{Erben}}, \citenamefont {{Hildebrandt}}, \citenamefont
  {{Hoekstra}}, \citenamefont {{Kitching}}, \citenamefont {{Mellier}},
  \citenamefont {{Simon}}, \citenamefont {{Bonnett}}, \citenamefont {{Coupon}},
  \citenamefont {{Fu}}, \citenamefont {{Harnois D{\'e}raps}}, \citenamefont
  {{Hudson}}, \citenamefont {{Kilbinger}}, \citenamefont {{Kuijken}},
  \citenamefont {{Rowe}}, \citenamefont {{Schrabback}}, \citenamefont
  {{Semboloni}}, \citenamefont {{van Uitert}}, \citenamefont {{Vafaei}},\ and\
  \citenamefont {{Velander}}}]{cfht1}%
  \BibitemOpen
  \bibfield  {author} {\bibinfo {author} {\bibfnamefont {C.}~\bibnamefont
  {{Heymans}}}, \bibinfo {author} {\bibfnamefont {L.}~\bibnamefont {{Van
  Waerbeke}}}, \bibinfo {author} {\bibfnamefont {L.}~\bibnamefont {{Miller}}},
  \bibinfo {author} {\bibfnamefont {T.}~\bibnamefont {{Erben}}}, \bibinfo
  {author} {\bibfnamefont {H.}~\bibnamefont {{Hildebrandt}}}, \bibinfo {author}
  {\bibfnamefont {H.}~\bibnamefont {{Hoekstra}}}, \bibinfo {author}
  {\bibfnamefont {T.~D.}\ \bibnamefont {{Kitching}}}, \bibinfo {author}
  {\bibfnamefont {Y.}~\bibnamefont {{Mellier}}}, \bibinfo {author}
  {\bibfnamefont {P.}~\bibnamefont {{Simon}}}, \bibinfo {author} {\bibfnamefont
  {C.}~\bibnamefont {{Bonnett}}}, \bibinfo {author} {\bibfnamefont
  {J.}~\bibnamefont {{Coupon}}}, \bibinfo {author} {\bibfnamefont
  {L.}~\bibnamefont {{Fu}}}, \bibinfo {author} {\bibfnamefont {J.}~\bibnamefont
  {{Harnois D{\'e}raps}}}, \bibinfo {author} {\bibfnamefont {M.~J.}\
  \bibnamefont {{Hudson}}}, \bibinfo {author} {\bibfnamefont {M.}~\bibnamefont
  {{Kilbinger}}}, \bibinfo {author} {\bibfnamefont {K.}~\bibnamefont
  {{Kuijken}}}, \bibinfo {author} {\bibfnamefont {B.}~\bibnamefont {{Rowe}}},
  \bibinfo {author} {\bibfnamefont {T.}~\bibnamefont {{Schrabback}}}, \bibinfo
  {author} {\bibfnamefont {E.}~\bibnamefont {{Semboloni}}}, \bibinfo {author}
  {\bibfnamefont {E.}~\bibnamefont {{van Uitert}}}, \bibinfo {author}
  {\bibfnamefont {S.}~\bibnamefont {{Vafaei}}}, \ and\ \bibinfo {author}
  {\bibfnamefont {M.}~\bibnamefont {{Velander}}},\ }\href {\doibase
  10.1111/j.1365-2966.2012.21952.x} {\bibfield  {journal} {\bibinfo  {journal}
  {\mnras}\ }\textbf {\bibinfo {volume} {427}},\ \bibinfo {pages} {146}
  (\bibinfo {year} {2012})},\ \Eprint {http://arxiv.org/abs/1210.0032}
  {arXiv:1210.0032 [astro-ph.CO]} \BibitemShut {NoStop}%
\bibitem [{\citenamefont {{Koekemoer}}\ \emph {et~al.}(2007)\citenamefont
  {{Koekemoer}}, \citenamefont {{Aussel}}, \citenamefont {{Calzetti}},
  \citenamefont {{Capak}}, \citenamefont {{Giavalisco}}, \citenamefont
  {{Kneib}}, \citenamefont {{Leauthaud}}, \citenamefont {{Le F{\`e}vre}},
  \citenamefont {{McCracken}}, \citenamefont {{Massey}}, \citenamefont
  {{Mobasher}}, \citenamefont {{Rhodes}}, \citenamefont {{Scoville}},\ and\
  \citenamefont {{Shopbell}}}]{cosmos}%
  \BibitemOpen
  \bibfield  {author} {\bibinfo {author} {\bibfnamefont {A.~M.}\ \bibnamefont
  {{Koekemoer}}}, \bibinfo {author} {\bibfnamefont {H.}~\bibnamefont
  {{Aussel}}}, \bibinfo {author} {\bibfnamefont {D.}~\bibnamefont
  {{Calzetti}}}, \bibinfo {author} {\bibfnamefont {P.}~\bibnamefont {{Capak}}},
  \bibinfo {author} {\bibfnamefont {M.}~\bibnamefont {{Giavalisco}}}, \bibinfo
  {author} {\bibfnamefont {J.-P.}\ \bibnamefont {{Kneib}}}, \bibinfo {author}
  {\bibfnamefont {A.}~\bibnamefont {{Leauthaud}}}, \bibinfo {author}
  {\bibfnamefont {O.}~\bibnamefont {{Le F{\`e}vre}}}, \bibinfo {author}
  {\bibfnamefont {H.~J.}\ \bibnamefont {{McCracken}}}, \bibinfo {author}
  {\bibfnamefont {R.}~\bibnamefont {{Massey}}}, \bibinfo {author}
  {\bibfnamefont {B.}~\bibnamefont {{Mobasher}}}, \bibinfo {author}
  {\bibfnamefont {J.}~\bibnamefont {{Rhodes}}}, \bibinfo {author}
  {\bibfnamefont {N.}~\bibnamefont {{Scoville}}}, \ and\ \bibinfo {author}
  {\bibfnamefont {P.~L.}\ \bibnamefont {{Shopbell}}},\ }\href {\doibase
  10.1086/520086} {\bibfield  {journal} {\bibinfo  {journal} {\apjs}\ }\textbf
  {\bibinfo {volume} {172}},\ \bibinfo {pages} {196} (\bibinfo {year}
  {2007})},\ \Eprint {http://arxiv.org/abs/astro-ph/0703095} {astro-ph/0703095}
  \BibitemShut {NoStop}%
\bibitem [{\citenamefont {{Gruen}}\ \emph {et~al.}(2016)\citenamefont
  {{Gruen}}, \citenamefont {{Friedrich}}, \citenamefont {{Amara}},
  \citenamefont {{Bacon}}, \citenamefont {{Bonnett}}, \citenamefont
  {{Hartley}}, \citenamefont {{Jain}}, \citenamefont {{Jarvis}}, \citenamefont
  {{Kacprzak}}, \citenamefont {{Krause}}, \citenamefont {{Mana}}, \citenamefont
  {{Rozo}}, \citenamefont {{Rykoff}}, \citenamefont {{Seitz}}, \citenamefont
  {{Sheldon}}, \citenamefont {{Troxel}}, \citenamefont {{Vikram}},
  \citenamefont {{Abbott}}, \citenamefont {{Abdalla}}, \citenamefont {{Allam}},
  \citenamefont {{Armstrong}}, \citenamefont {{Banerji}}, \citenamefont
  {{Bauer}}, \citenamefont {{Becker}}, \citenamefont {{Benoit-L{\'e}vy}},
  \citenamefont {{Bernstein}}, \citenamefont {{Bernstein}}, \citenamefont
  {{Bertin}}, \citenamefont {{Bridle}}, \citenamefont {{Brooks}}, \citenamefont
  {{Buckley-Geer}}, \citenamefont {{Burke}}, \citenamefont {{Capozzi}},
  \citenamefont {{Carnero Rosell}}, \citenamefont {{Carrasco Kind}},
  \citenamefont {{Carretero}}, \citenamefont {{Crocce}}, \citenamefont
  {{Cunha}}, \citenamefont {{D'Andrea}}, \citenamefont {{da Costa}},
  \citenamefont {{DePoy}}, \citenamefont {{Desai}}, \citenamefont {{Diehl}},
  \citenamefont {{Dietrich}}, \citenamefont {{Doel}}, \citenamefont {{Eifler}},
  \citenamefont {{Neto}}, \citenamefont {{Fernandez}}, \citenamefont
  {{Flaugher}}, \citenamefont {{Fosalba}}, \citenamefont {{Frieman}},
  \citenamefont {{Gerdes}}, \citenamefont {{Gruendl}}, \citenamefont
  {{Gutierrez}}, \citenamefont {{Honscheid}}, \citenamefont {{James}},
  \citenamefont {{Kuehn}}, \citenamefont {{Kuropatkin}}, \citenamefont
  {{Lahav}}, \citenamefont {{Li}}, \citenamefont {{Lima}}, \citenamefont
  {{Maia}}, \citenamefont {{March}}, \citenamefont {{Martini}}, \citenamefont
  {{Melchior}}, \citenamefont {{Miller}}, \citenamefont {{Miquel}},
  \citenamefont {{Mohr}}, \citenamefont {{Nord}}, \citenamefont {{Ogando}},
  \citenamefont {{Plazas}}, \citenamefont {{Reil}}, \citenamefont {{Romer}},
  \citenamefont {{Roodman}}, \citenamefont {{Sako}}, \citenamefont {{Sanchez}},
  \citenamefont {{Scarpine}}, \citenamefont {{Schubnell}}, \citenamefont
  {{Sevilla-Noarbe}}, \citenamefont {{Smith}}, \citenamefont {{Soares-Santos}},
  \citenamefont {{Sobreira}}, \citenamefont {{Suchyta}}, \citenamefont
  {{Swanson}}, \citenamefont {{Tarle}}, \citenamefont {{Thaler}}, \citenamefont
  {{Thomas}}, \citenamefont {{Walker}}, \citenamefont {{Wechsler}},
  \citenamefont {{Weller}}, \citenamefont {{Zhang}},\ and\ \citenamefont
  {{Zuntz}}}]{DES}%
  \BibitemOpen
  \bibfield  {author} {\bibinfo {author} {\bibfnamefont {D.}~\bibnamefont
  {{Gruen}}}, \bibinfo {author} {\bibfnamefont {O.}~\bibnamefont
  {{Friedrich}}}, \bibinfo {author} {\bibfnamefont {A.}~\bibnamefont
  {{Amara}}}, \bibinfo {author} {\bibfnamefont {D.}~\bibnamefont {{Bacon}}},
  \bibinfo {author} {\bibfnamefont {C.}~\bibnamefont {{Bonnett}}}, \bibinfo
  {author} {\bibfnamefont {W.}~\bibnamefont {{Hartley}}}, \bibinfo {author}
  {\bibfnamefont {B.}~\bibnamefont {{Jain}}}, \bibinfo {author} {\bibfnamefont
  {M.}~\bibnamefont {{Jarvis}}}, \bibinfo {author} {\bibfnamefont
  {T.}~\bibnamefont {{Kacprzak}}}, \bibinfo {author} {\bibfnamefont
  {E.}~\bibnamefont {{Krause}}}, \bibinfo {author} {\bibfnamefont
  {A.}~\bibnamefont {{Mana}}}, \bibinfo {author} {\bibfnamefont
  {E.}~\bibnamefont {{Rozo}}}, \bibinfo {author} {\bibfnamefont {E.~S.}\
  \bibnamefont {{Rykoff}}}, \bibinfo {author} {\bibfnamefont {S.}~\bibnamefont
  {{Seitz}}}, \bibinfo {author} {\bibfnamefont {E.}~\bibnamefont {{Sheldon}}},
  \bibinfo {author} {\bibfnamefont {M.~A.}\ \bibnamefont {{Troxel}}}, \bibinfo
  {author} {\bibfnamefont {V.}~\bibnamefont {{Vikram}}}, \bibinfo {author}
  {\bibfnamefont {T.~M.~C.}\ \bibnamefont {{Abbott}}}, \bibinfo {author}
  {\bibfnamefont {F.~B.}\ \bibnamefont {{Abdalla}}}, \bibinfo {author}
  {\bibfnamefont {S.}~\bibnamefont {{Allam}}}, \bibinfo {author} {\bibfnamefont
  {R.}~\bibnamefont {{Armstrong}}}, \bibinfo {author} {\bibfnamefont
  {M.}~\bibnamefont {{Banerji}}}, \bibinfo {author} {\bibfnamefont {A.~H.}\
  \bibnamefont {{Bauer}}}, \bibinfo {author} {\bibfnamefont {M.~R.}\
  \bibnamefont {{Becker}}}, \bibinfo {author} {\bibfnamefont {A.}~\bibnamefont
  {{Benoit-L{\'e}vy}}}, \bibinfo {author} {\bibfnamefont {G.~M.}\ \bibnamefont
  {{Bernstein}}}, \bibinfo {author} {\bibfnamefont {R.~A.}\ \bibnamefont
  {{Bernstein}}}, \bibinfo {author} {\bibfnamefont {E.}~\bibnamefont
  {{Bertin}}}, \bibinfo {author} {\bibfnamefont {S.~L.}\ \bibnamefont
  {{Bridle}}}, \bibinfo {author} {\bibfnamefont {D.}~\bibnamefont {{Brooks}}},
  \bibinfo {author} {\bibfnamefont {E.}~\bibnamefont {{Buckley-Geer}}},
  \bibinfo {author} {\bibfnamefont {D.~L.}\ \bibnamefont {{Burke}}}, \bibinfo
  {author} {\bibfnamefont {D.}~\bibnamefont {{Capozzi}}}, \bibinfo {author}
  {\bibfnamefont {A.}~\bibnamefont {{Carnero Rosell}}}, \bibinfo {author}
  {\bibfnamefont {M.}~\bibnamefont {{Carrasco Kind}}}, \bibinfo {author}
  {\bibfnamefont {J.}~\bibnamefont {{Carretero}}}, \bibinfo {author}
  {\bibfnamefont {M.}~\bibnamefont {{Crocce}}}, \bibinfo {author}
  {\bibfnamefont {C.~E.}\ \bibnamefont {{Cunha}}}, \bibinfo {author}
  {\bibfnamefont {C.~B.}\ \bibnamefont {{D'Andrea}}}, \bibinfo {author}
  {\bibfnamefont {L.~N.}\ \bibnamefont {{da Costa}}}, \bibinfo {author}
  {\bibfnamefont {D.~L.}\ \bibnamefont {{DePoy}}}, \bibinfo {author}
  {\bibfnamefont {S.}~\bibnamefont {{Desai}}}, \bibinfo {author} {\bibfnamefont
  {H.~T.}\ \bibnamefont {{Diehl}}}, \bibinfo {author} {\bibfnamefont {J.~P.}\
  \bibnamefont {{Dietrich}}}, \bibinfo {author} {\bibfnamefont
  {P.}~\bibnamefont {{Doel}}}, \bibinfo {author} {\bibfnamefont {T.~F.}\
  \bibnamefont {{Eifler}}}, \bibinfo {author} {\bibfnamefont {A.~F.}\
  \bibnamefont {{Neto}}}, \bibinfo {author} {\bibfnamefont {E.}~\bibnamefont
  {{Fernandez}}}, \bibinfo {author} {\bibfnamefont {B.}~\bibnamefont
  {{Flaugher}}}, \bibinfo {author} {\bibfnamefont {P.}~\bibnamefont
  {{Fosalba}}}, \bibinfo {author} {\bibfnamefont {J.}~\bibnamefont
  {{Frieman}}}, \bibinfo {author} {\bibfnamefont {D.~W.}\ \bibnamefont
  {{Gerdes}}}, \bibinfo {author} {\bibfnamefont {R.~A.}\ \bibnamefont
  {{Gruendl}}}, \bibinfo {author} {\bibfnamefont {G.}~\bibnamefont
  {{Gutierrez}}}, \bibinfo {author} {\bibfnamefont {K.}~\bibnamefont
  {{Honscheid}}}, \bibinfo {author} {\bibfnamefont {D.~J.}\ \bibnamefont
  {{James}}}, \bibinfo {author} {\bibfnamefont {K.}~\bibnamefont {{Kuehn}}},
  \bibinfo {author} {\bibfnamefont {N.}~\bibnamefont {{Kuropatkin}}}, \bibinfo
  {author} {\bibfnamefont {O.}~\bibnamefont {{Lahav}}}, \bibinfo {author}
  {\bibfnamefont {T.~S.}\ \bibnamefont {{Li}}}, \bibinfo {author}
  {\bibfnamefont {M.}~\bibnamefont {{Lima}}}, \bibinfo {author} {\bibfnamefont
  {M.~A.~G.}\ \bibnamefont {{Maia}}}, \bibinfo {author} {\bibfnamefont
  {M.}~\bibnamefont {{March}}}, \bibinfo {author} {\bibfnamefont
  {P.}~\bibnamefont {{Martini}}}, \bibinfo {author} {\bibfnamefont
  {P.}~\bibnamefont {{Melchior}}}, \bibinfo {author} {\bibfnamefont {C.~J.}\
  \bibnamefont {{Miller}}}, \bibinfo {author} {\bibfnamefont {R.}~\bibnamefont
  {{Miquel}}}, \bibinfo {author} {\bibfnamefont {J.~J.}\ \bibnamefont
  {{Mohr}}}, \bibinfo {author} {\bibfnamefont {B.}~\bibnamefont {{Nord}}},
  \bibinfo {author} {\bibfnamefont {R.}~\bibnamefont {{Ogando}}}, \bibinfo
  {author} {\bibfnamefont {A.~A.}\ \bibnamefont {{Plazas}}}, \bibinfo {author}
  {\bibfnamefont {K.}~\bibnamefont {{Reil}}}, \bibinfo {author} {\bibfnamefont
  {A.~K.}\ \bibnamefont {{Romer}}}, \bibinfo {author} {\bibfnamefont
  {A.}~\bibnamefont {{Roodman}}}, \bibinfo {author} {\bibfnamefont
  {M.}~\bibnamefont {{Sako}}}, \bibinfo {author} {\bibfnamefont
  {E.}~\bibnamefont {{Sanchez}}}, \bibinfo {author} {\bibfnamefont
  {V.}~\bibnamefont {{Scarpine}}}, \bibinfo {author} {\bibfnamefont
  {M.}~\bibnamefont {{Schubnell}}}, \bibinfo {author} {\bibfnamefont
  {I.}~\bibnamefont {{Sevilla-Noarbe}}}, \bibinfo {author} {\bibfnamefont
  {R.~C.}\ \bibnamefont {{Smith}}}, \bibinfo {author} {\bibfnamefont
  {M.}~\bibnamefont {{Soares-Santos}}}, \bibinfo {author} {\bibfnamefont
  {F.}~\bibnamefont {{Sobreira}}}, \bibinfo {author} {\bibfnamefont
  {E.}~\bibnamefont {{Suchyta}}}, \bibinfo {author} {\bibfnamefont {M.~E.~C.}\
  \bibnamefont {{Swanson}}}, \bibinfo {author} {\bibfnamefont {G.}~\bibnamefont
  {{Tarle}}}, \bibinfo {author} {\bibfnamefont {J.}~\bibnamefont {{Thaler}}},
  \bibinfo {author} {\bibfnamefont {D.}~\bibnamefont {{Thomas}}}, \bibinfo
  {author} {\bibfnamefont {A.~R.}\ \bibnamefont {{Walker}}}, \bibinfo {author}
  {\bibfnamefont {R.~H.}\ \bibnamefont {{Wechsler}}}, \bibinfo {author}
  {\bibfnamefont {J.}~\bibnamefont {{Weller}}}, \bibinfo {author}
  {\bibfnamefont {Y.}~\bibnamefont {{Zhang}}}, \ and\ \bibinfo {author}
  {\bibfnamefont {J.}~\bibnamefont {{Zuntz}}},\ }\href {\doibase
  10.1093/mnras/stv2506} {\bibfield  {journal} {\bibinfo  {journal} {\mnras}\
  }\textbf {\bibinfo {volume} {455}},\ \bibinfo {pages} {3367} (\bibinfo {year}
  {2016})},\ \Eprint {http://arxiv.org/abs/1507.05090} {arXiv:1507.05090}
  \BibitemShut {NoStop}%
\bibitem [{\citenamefont {{LSST Dark Energy Science
  Collaboration}}(2012)}]{LSST}%
  \BibitemOpen
  \bibfield  {author} {\bibinfo {author} {\bibnamefont {{LSST Dark Energy
  Science Collaboration}}},\ }\href@noop {} {\bibfield  {journal} {\bibinfo
  {journal} {ArXiv e-prints}\ } (\bibinfo {year} {2012})},\ \Eprint
  {http://arxiv.org/abs/1211.0310} {arXiv:1211.0310 [astro-ph.CO]} \BibitemShut
  {NoStop}%
\bibitem [{\citenamefont {{Spergel}}\ \emph {et~al.}(2015)\citenamefont
  {{Spergel}}, \citenamefont {{Gehrels}}, \citenamefont {{Baltay}},
  \citenamefont {{Bennett}}, \citenamefont {{Breckinridge}}, \citenamefont
  {{Donahue}}, \citenamefont {{Dressler}}, \citenamefont {{Gaudi}},
  \citenamefont {{Greene}}, \citenamefont {{Guyon}}, \citenamefont {{Hirata}},
  \citenamefont {{Kalirai}}, \citenamefont {{Kasdin}}, \citenamefont
  {{Macintosh}}, \citenamefont {{Moos}}, \citenamefont {{Perlmutter}},
  \citenamefont {{Postman}}, \citenamefont {{Rauscher}}, \citenamefont
  {{Rhodes}}, \citenamefont {{Wang}}, \citenamefont {{Weinberg}}, \citenamefont
  {{Benford}}, \citenamefont {{Hudson}}, \citenamefont {{Jeong}}, \citenamefont
  {{Mellier}}, \citenamefont {{Traub}}, \citenamefont {{Yamada}}, \citenamefont
  {{Capak}}, \citenamefont {{Colbert}}, \citenamefont {{Masters}},
  \citenamefont {{Penny}}, \citenamefont {{Savransky}}, \citenamefont
  {{Stern}}, \citenamefont {{Zimmerman}}, \citenamefont {{Barry}},
  \citenamefont {{Bartusek}}, \citenamefont {{Carpenter}}, \citenamefont
  {{Cheng}}, \citenamefont {{Content}}, \citenamefont {{Dekens}}, \citenamefont
  {{Demers}}, \citenamefont {{Grady}}, \citenamefont {{Jackson}}, \citenamefont
  {{Kuan}}, \citenamefont {{Kruk}}, \citenamefont {{Melton}}, \citenamefont
  {{Nemati}}, \citenamefont {{Parvin}}, \citenamefont {{Poberezhskiy}},
  \citenamefont {{Peddie}}, \citenamefont {{Ruffa}}, \citenamefont {{Wallace}},
  \citenamefont {{Whipple}}, \citenamefont {{Wollack}},\ and\ \citenamefont
  {{Zhao}}}]{WFIRST}%
  \BibitemOpen
  \bibfield  {author} {\bibinfo {author} {\bibfnamefont {D.}~\bibnamefont
  {{Spergel}}}, \bibinfo {author} {\bibfnamefont {N.}~\bibnamefont
  {{Gehrels}}}, \bibinfo {author} {\bibfnamefont {C.}~\bibnamefont {{Baltay}}},
  \bibinfo {author} {\bibfnamefont {D.}~\bibnamefont {{Bennett}}}, \bibinfo
  {author} {\bibfnamefont {J.}~\bibnamefont {{Breckinridge}}}, \bibinfo
  {author} {\bibfnamefont {M.}~\bibnamefont {{Donahue}}}, \bibinfo {author}
  {\bibfnamefont {A.}~\bibnamefont {{Dressler}}}, \bibinfo {author}
  {\bibfnamefont {B.~S.}\ \bibnamefont {{Gaudi}}}, \bibinfo {author}
  {\bibfnamefont {T.}~\bibnamefont {{Greene}}}, \bibinfo {author}
  {\bibfnamefont {O.}~\bibnamefont {{Guyon}}}, \bibinfo {author} {\bibfnamefont
  {C.}~\bibnamefont {{Hirata}}}, \bibinfo {author} {\bibfnamefont
  {J.}~\bibnamefont {{Kalirai}}}, \bibinfo {author} {\bibfnamefont {N.~J.}\
  \bibnamefont {{Kasdin}}}, \bibinfo {author} {\bibfnamefont {B.}~\bibnamefont
  {{Macintosh}}}, \bibinfo {author} {\bibfnamefont {W.}~\bibnamefont {{Moos}}},
  \bibinfo {author} {\bibfnamefont {S.}~\bibnamefont {{Perlmutter}}}, \bibinfo
  {author} {\bibfnamefont {M.}~\bibnamefont {{Postman}}}, \bibinfo {author}
  {\bibfnamefont {B.}~\bibnamefont {{Rauscher}}}, \bibinfo {author}
  {\bibfnamefont {J.}~\bibnamefont {{Rhodes}}}, \bibinfo {author}
  {\bibfnamefont {Y.}~\bibnamefont {{Wang}}}, \bibinfo {author} {\bibfnamefont
  {D.}~\bibnamefont {{Weinberg}}}, \bibinfo {author} {\bibfnamefont
  {D.}~\bibnamefont {{Benford}}}, \bibinfo {author} {\bibfnamefont
  {M.}~\bibnamefont {{Hudson}}}, \bibinfo {author} {\bibfnamefont {W.-S.}\
  \bibnamefont {{Jeong}}}, \bibinfo {author} {\bibfnamefont {Y.}~\bibnamefont
  {{Mellier}}}, \bibinfo {author} {\bibfnamefont {W.}~\bibnamefont {{Traub}}},
  \bibinfo {author} {\bibfnamefont {T.}~\bibnamefont {{Yamada}}}, \bibinfo
  {author} {\bibfnamefont {P.}~\bibnamefont {{Capak}}}, \bibinfo {author}
  {\bibfnamefont {J.}~\bibnamefont {{Colbert}}}, \bibinfo {author}
  {\bibfnamefont {D.}~\bibnamefont {{Masters}}}, \bibinfo {author}
  {\bibfnamefont {M.}~\bibnamefont {{Penny}}}, \bibinfo {author} {\bibfnamefont
  {D.}~\bibnamefont {{Savransky}}}, \bibinfo {author} {\bibfnamefont
  {D.}~\bibnamefont {{Stern}}}, \bibinfo {author} {\bibfnamefont
  {N.}~\bibnamefont {{Zimmerman}}}, \bibinfo {author} {\bibfnamefont
  {R.}~\bibnamefont {{Barry}}}, \bibinfo {author} {\bibfnamefont
  {L.}~\bibnamefont {{Bartusek}}}, \bibinfo {author} {\bibfnamefont
  {K.}~\bibnamefont {{Carpenter}}}, \bibinfo {author} {\bibfnamefont
  {E.}~\bibnamefont {{Cheng}}}, \bibinfo {author} {\bibfnamefont
  {D.}~\bibnamefont {{Content}}}, \bibinfo {author} {\bibfnamefont
  {F.}~\bibnamefont {{Dekens}}}, \bibinfo {author} {\bibfnamefont
  {R.}~\bibnamefont {{Demers}}}, \bibinfo {author} {\bibfnamefont
  {K.}~\bibnamefont {{Grady}}}, \bibinfo {author} {\bibfnamefont
  {C.}~\bibnamefont {{Jackson}}}, \bibinfo {author} {\bibfnamefont
  {G.}~\bibnamefont {{Kuan}}}, \bibinfo {author} {\bibfnamefont
  {J.}~\bibnamefont {{Kruk}}}, \bibinfo {author} {\bibfnamefont
  {M.}~\bibnamefont {{Melton}}}, \bibinfo {author} {\bibfnamefont
  {B.}~\bibnamefont {{Nemati}}}, \bibinfo {author} {\bibfnamefont
  {B.}~\bibnamefont {{Parvin}}}, \bibinfo {author} {\bibfnamefont
  {I.}~\bibnamefont {{Poberezhskiy}}}, \bibinfo {author} {\bibfnamefont
  {C.}~\bibnamefont {{Peddie}}}, \bibinfo {author} {\bibfnamefont
  {J.}~\bibnamefont {{Ruffa}}}, \bibinfo {author} {\bibfnamefont {J.~K.}\
  \bibnamefont {{Wallace}}}, \bibinfo {author} {\bibfnamefont {A.}~\bibnamefont
  {{Whipple}}}, \bibinfo {author} {\bibfnamefont {E.}~\bibnamefont
  {{Wollack}}}, \ and\ \bibinfo {author} {\bibfnamefont {F.}~\bibnamefont
  {{Zhao}}},\ }\href@noop {} {\bibfield  {journal} {\bibinfo  {journal} {ArXiv
  e-prints}\ } (\bibinfo {year} {2015})},\ \Eprint
  {http://arxiv.org/abs/1503.03757} {arXiv:1503.03757 [astro-ph.IM]}
  \BibitemShut {NoStop}%
\bibitem [{\citenamefont {{Amiaux}}\ \emph {et~al.}(2012)\citenamefont
  {{Amiaux}}, \citenamefont {{Scaramella}}, \citenamefont {{Mellier}},
  \citenamefont {{Altieri}}, \citenamefont {{Burigana}}, \citenamefont {{Da
  Silva}}, \citenamefont {{Gomez}}, \citenamefont {{Hoar}}, \citenamefont
  {{Laureijs}}, \citenamefont {{Maiorano}}, \citenamefont {{Magalh{\~a}es
  Oliveira}}, \citenamefont {{Renk}}, \citenamefont {{Saavedra Criado}},
  \citenamefont {{Tereno}}, \citenamefont {{Augu{\`e}res}}, \citenamefont
  {{Brinchmann}}, \citenamefont {{Cropper}}, \citenamefont {{Duvet}},
  \citenamefont {{Ealet}}, \citenamefont {{Franzetti}}, \citenamefont
  {{Garilli}}, \citenamefont {{Gondoin}}, \citenamefont {{Guzzo}},
  \citenamefont {{Hoekstra}}, \citenamefont {{Holmes}}, \citenamefont
  {{Jahnke}}, \citenamefont {{Kitching}}, \citenamefont {{Meneghetti}},
  \citenamefont {{Percival}},\ and\ \citenamefont {{Warren}}}]{Euclid}%
  \BibitemOpen
  \bibfield  {author} {\bibinfo {author} {\bibfnamefont {J.}~\bibnamefont
  {{Amiaux}}}, \bibinfo {author} {\bibfnamefont {R.}~\bibnamefont
  {{Scaramella}}}, \bibinfo {author} {\bibfnamefont {Y.}~\bibnamefont
  {{Mellier}}}, \bibinfo {author} {\bibfnamefont {B.}~\bibnamefont
  {{Altieri}}}, \bibinfo {author} {\bibfnamefont {C.}~\bibnamefont
  {{Burigana}}}, \bibinfo {author} {\bibfnamefont {A.}~\bibnamefont {{Da
  Silva}}}, \bibinfo {author} {\bibfnamefont {P.}~\bibnamefont {{Gomez}}},
  \bibinfo {author} {\bibfnamefont {J.}~\bibnamefont {{Hoar}}}, \bibinfo
  {author} {\bibfnamefont {R.}~\bibnamefont {{Laureijs}}}, \bibinfo {author}
  {\bibfnamefont {E.}~\bibnamefont {{Maiorano}}}, \bibinfo {author}
  {\bibfnamefont {D.}~\bibnamefont {{Magalh{\~a}es Oliveira}}}, \bibinfo
  {author} {\bibfnamefont {F.}~\bibnamefont {{Renk}}}, \bibinfo {author}
  {\bibfnamefont {G.}~\bibnamefont {{Saavedra Criado}}}, \bibinfo {author}
  {\bibfnamefont {I.}~\bibnamefont {{Tereno}}}, \bibinfo {author}
  {\bibfnamefont {J.~L.}\ \bibnamefont {{Augu{\`e}res}}}, \bibinfo {author}
  {\bibfnamefont {J.}~\bibnamefont {{Brinchmann}}}, \bibinfo {author}
  {\bibfnamefont {M.}~\bibnamefont {{Cropper}}}, \bibinfo {author}
  {\bibfnamefont {L.}~\bibnamefont {{Duvet}}}, \bibinfo {author} {\bibfnamefont
  {A.}~\bibnamefont {{Ealet}}}, \bibinfo {author} {\bibfnamefont
  {P.}~\bibnamefont {{Franzetti}}}, \bibinfo {author} {\bibfnamefont
  {B.}~\bibnamefont {{Garilli}}}, \bibinfo {author} {\bibfnamefont
  {P.}~\bibnamefont {{Gondoin}}}, \bibinfo {author} {\bibfnamefont
  {L.}~\bibnamefont {{Guzzo}}}, \bibinfo {author} {\bibfnamefont
  {H.}~\bibnamefont {{Hoekstra}}}, \bibinfo {author} {\bibfnamefont
  {R.}~\bibnamefont {{Holmes}}}, \bibinfo {author} {\bibfnamefont
  {K.}~\bibnamefont {{Jahnke}}}, \bibinfo {author} {\bibfnamefont
  {T.}~\bibnamefont {{Kitching}}}, \bibinfo {author} {\bibfnamefont
  {M.}~\bibnamefont {{Meneghetti}}}, \bibinfo {author} {\bibfnamefont
  {W.}~\bibnamefont {{Percival}}}, \ and\ \bibinfo {author} {\bibfnamefont
  {S.}~\bibnamefont {{Warren}}},\ }in\ \href {\doibase 10.1117/12.926513}
  {\emph {\bibinfo {booktitle} {Society of Photo-Optical Instrumentation
  Engineers (SPIE) Conference Series}}},\ \bibinfo {series} {Society of
  Photo-Optical Instrumentation Engineers (SPIE) Conference Series}, Vol.\
  \bibinfo {volume} {8442}\ (\bibinfo {year} {2012})\ p.\ \bibinfo {pages}
  {84420Z},\ \Eprint {http://arxiv.org/abs/1209.2228} {arXiv:1209.2228
  [astro-ph.IM]} \BibitemShut {NoStop}%
\bibitem [{\citenamefont {{Kratochvil}}\ \emph {et~al.}(2012)\citenamefont
  {{Kratochvil}}, \citenamefont {{Lim}}, \citenamefont {{Wang}}, \citenamefont
  {{Haiman}}, \citenamefont {{May}},\ and\ \citenamefont
  {{Huffenberger}}}]{MinkJan}%
  \BibitemOpen
  \bibfield  {author} {\bibinfo {author} {\bibfnamefont {J.~M.}\ \bibnamefont
  {{Kratochvil}}}, \bibinfo {author} {\bibfnamefont {E.~A.}\ \bibnamefont
  {{Lim}}}, \bibinfo {author} {\bibfnamefont {S.}~\bibnamefont {{Wang}}},
  \bibinfo {author} {\bibfnamefont {Z.}~\bibnamefont {{Haiman}}}, \bibinfo
  {author} {\bibfnamefont {M.}~\bibnamefont {{May}}}, \ and\ \bibinfo {author}
  {\bibfnamefont {K.}~\bibnamefont {{Huffenberger}}},\ }\href {\doibase
  10.1103/PhysRevD.85.103513} {\bibfield  {journal} {\bibinfo  {journal}
  {\prd}\ }\textbf {\bibinfo {volume} {85}},\ \bibinfo {eid} {103513} (\bibinfo
  {year} {2012})},\ \Eprint {http://arxiv.org/abs/1109.6334} {arXiv:1109.6334
  [astro-ph.CO]} \BibitemShut {NoStop}%
\bibitem [{\citenamefont {{Yang}}\ \emph {et~al.}(2011)\citenamefont {{Yang}},
  \citenamefont {{Kratochvil}}, \citenamefont {{Wang}}, \citenamefont {{Lim}},
  \citenamefont {{Haiman}},\ and\ \citenamefont {{May}}}]{PeaksJan}%
  \BibitemOpen
  \bibfield  {author} {\bibinfo {author} {\bibfnamefont {X.}~\bibnamefont
  {{Yang}}}, \bibinfo {author} {\bibfnamefont {J.~M.}\ \bibnamefont
  {{Kratochvil}}}, \bibinfo {author} {\bibfnamefont {S.}~\bibnamefont
  {{Wang}}}, \bibinfo {author} {\bibfnamefont {E.~A.}\ \bibnamefont {{Lim}}},
  \bibinfo {author} {\bibfnamefont {Z.}~\bibnamefont {{Haiman}}}, \ and\
  \bibinfo {author} {\bibfnamefont {M.}~\bibnamefont {{May}}},\ }\href
  {\doibase 10.1103/PhysRevD.84.043529} {\bibfield  {journal} {\bibinfo
  {journal} {\prd}\ }\textbf {\bibinfo {volume} {84}},\ \bibinfo {eid} {043529}
  (\bibinfo {year} {2011})},\ \Eprint {http://arxiv.org/abs/1109.6333}
  {arXiv:1109.6333} \BibitemShut {NoStop}%
\bibitem [{\citenamefont {{Marian}}\ \emph {et~al.}(2009)\citenamefont
  {{Marian}}, \citenamefont {{Smith}},\ and\ \citenamefont
  {{Bernstein}}}]{NG-Marian}%
  \BibitemOpen
  \bibfield  {author} {\bibinfo {author} {\bibfnamefont {L.}~\bibnamefont
  {{Marian}}}, \bibinfo {author} {\bibfnamefont {R.~E.}\ \bibnamefont
  {{Smith}}}, \ and\ \bibinfo {author} {\bibfnamefont {G.~M.}\ \bibnamefont
  {{Bernstein}}},\ }\href {\doibase 10.1088/0004-637X/698/1/L33} {\bibfield
  {journal} {\bibinfo  {journal} {\apjl}\ }\textbf {\bibinfo {volume} {698}},\
  \bibinfo {pages} {L33} (\bibinfo {year} {2009})},\ \Eprint
  {http://arxiv.org/abs/0811.1991} {arXiv:0811.1991} \BibitemShut {NoStop}%
\bibitem [{\citenamefont {{Takada}}\ and\ \citenamefont
  {{Jain}}(2002)}]{NG-Jain1}%
  \BibitemOpen
  \bibfield  {author} {\bibinfo {author} {\bibfnamefont {M.}~\bibnamefont
  {{Takada}}}\ and\ \bibinfo {author} {\bibfnamefont {B.}~\bibnamefont
  {{Jain}}},\ }\href {\doibase 10.1046/j.1365-8711.2002.05972.x} {\bibfield
  {journal} {\bibinfo  {journal} {\mnras}\ }\textbf {\bibinfo {volume} {337}},\
  \bibinfo {pages} {875} (\bibinfo {year} {2002})},\ \Eprint
  {http://arxiv.org/abs/astro-ph/0205055} {astro-ph/0205055} \BibitemShut
  {NoStop}%
\bibitem [{\citenamefont {{Takada}}\ and\ \citenamefont
  {{Jain}}(2003)}]{NG-Jain2}%
  \BibitemOpen
  \bibfield  {author} {\bibinfo {author} {\bibfnamefont {M.}~\bibnamefont
  {{Takada}}}\ and\ \bibinfo {author} {\bibfnamefont {B.}~\bibnamefont
  {{Jain}}},\ }\href {\doibase 10.1046/j.1365-8711.2003.06868.x} {\bibfield
  {journal} {\bibinfo  {journal} {\mnras}\ }\textbf {\bibinfo {volume} {344}},\
  \bibinfo {pages} {857} (\bibinfo {year} {2003})},\ \Eprint
  {http://arxiv.org/abs/astro-ph/0304034} {astro-ph/0304034} \BibitemShut
  {NoStop}%
\bibitem [{\citenamefont {{Takada}}\ and\ \citenamefont
  {{Jain}}(2004)}]{NG-Jain3}%
  \BibitemOpen
  \bibfield  {author} {\bibinfo {author} {\bibfnamefont {M.}~\bibnamefont
  {{Takada}}}\ and\ \bibinfo {author} {\bibfnamefont {B.}~\bibnamefont
  {{Jain}}},\ }\href {\doibase 10.1111/j.1365-2966.2004.07410.x} {\bibfield
  {journal} {\bibinfo  {journal} {\mnras}\ }\textbf {\bibinfo {volume} {348}},\
  \bibinfo {pages} {897} (\bibinfo {year} {2004})},\ \Eprint
  {http://arxiv.org/abs/astro-ph/0310125} {astro-ph/0310125} \BibitemShut
  {NoStop}%
\bibitem [{\citenamefont {{Berg{\'e}}}\ \emph {et~al.}(2010)\citenamefont
  {{Berg{\'e}}}, \citenamefont {{Amara}},\ and\ \citenamefont
  {{R{\'e}fr{\'e}gier}}}]{NG-Refregier}%
  \BibitemOpen
  \bibfield  {author} {\bibinfo {author} {\bibfnamefont {J.}~\bibnamefont
  {{Berg{\'e}}}}, \bibinfo {author} {\bibfnamefont {A.}~\bibnamefont
  {{Amara}}}, \ and\ \bibinfo {author} {\bibfnamefont {A.}~\bibnamefont
  {{R{\'e}fr{\'e}gier}}},\ }\href {\doibase 10.1088/0004-637X/712/2/992}
  {\bibfield  {journal} {\bibinfo  {journal} {\apj}\ }\textbf {\bibinfo
  {volume} {712}},\ \bibinfo {pages} {992} (\bibinfo {year} {2010})},\ \Eprint
  {http://arxiv.org/abs/0909.0529} {arXiv:0909.0529} \BibitemShut {NoStop}%
\bibitem [{\citenamefont {{Dietrich}}\ and\ \citenamefont
  {{Hartlap}}(2010)}]{NG-Dietrich}%
  \BibitemOpen
  \bibfield  {author} {\bibinfo {author} {\bibfnamefont {J.~P.}\ \bibnamefont
  {{Dietrich}}}\ and\ \bibinfo {author} {\bibfnamefont {J.}~\bibnamefont
  {{Hartlap}}},\ }\href {\doibase 10.1111/j.1365-2966.2009.15948.x} {\bibfield
  {journal} {\bibinfo  {journal} {\mnras}\ }\textbf {\bibinfo {volume} {402}},\
  \bibinfo {pages} {1049} (\bibinfo {year} {2010})},\ \Eprint
  {http://arxiv.org/abs/0906.3512} {arXiv:0906.3512} \BibitemShut {NoStop}%
\bibitem [{\citenamefont {{Song}}\ and\ \citenamefont
  {{Knox}}(2004)}]{SongKnox}%
  \BibitemOpen
  \bibfield  {author} {\bibinfo {author} {\bibfnamefont {Y.-S.}\ \bibnamefont
  {{Song}}}\ and\ \bibinfo {author} {\bibfnamefont {L.}~\bibnamefont
  {{Knox}}},\ }\href {\doibase 10.1103/PhysRevD.70.063510} {\bibfield
  {journal} {\bibinfo  {journal} {\prd}\ }\textbf {\bibinfo {volume} {70}},\
  \bibinfo {eid} {063510} (\bibinfo {year} {2004})},\ \Eprint
  {http://arxiv.org/abs/arXiv:astro-ph/0312175} {arXiv:astro-ph/0312175}
  \BibitemShut {NoStop}%
\bibitem [{\citenamefont {{Fang}}\ and\ \citenamefont
  {{Haiman}}(2007)}]{FangHaiman07}%
  \BibitemOpen
  \bibfield  {author} {\bibinfo {author} {\bibfnamefont {W.}~\bibnamefont
  {{Fang}}}\ and\ \bibinfo {author} {\bibfnamefont {Z.}~\bibnamefont
  {{Haiman}}},\ }\href {\doibase 10.1103/PhysRevD.75.043010} {\bibfield
  {journal} {\bibinfo  {journal} {\prd}\ }\textbf {\bibinfo {volume} {75}},\
  \bibinfo {eid} {043010} (\bibinfo {year} {2007})},\ \Eprint
  {http://arxiv.org/abs/astro-ph/0612187} {astro-ph/0612187} \BibitemShut
  {NoStop}%
\bibitem [{\citenamefont {{{Huterer}, D. and {Takada}, M. and {Bernstein}, G.
  and {Jain}, B.}}(2006)}]{Huterer2006}%
  \BibitemOpen
  \bibfield  {author} {\bibinfo {author} {\bibnamefont {{{Huterer}, D. and
  {Takada}, M. and {Bernstein}, G. and {Jain}, B.}}},\ }\href {\doibase
  10.1111/j.1365-2966.2005.09782.x} {\bibfield  {journal} {\bibinfo  {journal}
  {\mnras}\ }\textbf {\bibinfo {volume} {366}},\ \bibinfo {pages} {101}
  (\bibinfo {year} {2006})},\ \Eprint {http://arxiv.org/abs/astro-ph/0506030}
  {astro-ph/0506030} \BibitemShut {NoStop}%
\bibitem [{\citenamefont {{Martinet}}\ \emph {et~al.}(2015)\citenamefont
  {{Martinet}}, \citenamefont {{Bartlett}}, \citenamefont {{Kiessling}},\ and\
  \citenamefont {{Sartoris}}}]{MartinetPeaksTomo}%
  \BibitemOpen
  \bibfield  {author} {\bibinfo {author} {\bibfnamefont {N.}~\bibnamefont
  {{Martinet}}}, \bibinfo {author} {\bibfnamefont {J.~G.}\ \bibnamefont
  {{Bartlett}}}, \bibinfo {author} {\bibfnamefont {A.}~\bibnamefont
  {{Kiessling}}}, \ and\ \bibinfo {author} {\bibfnamefont {B.}~\bibnamefont
  {{Sartoris}}},\ }\href {\doibase 10.1051/0004-6361/201425164} {\bibfield
  {journal} {\bibinfo  {journal} {\aap}\ }\textbf {\bibinfo {volume} {581}},\
  \bibinfo {eid} {A101} (\bibinfo {year} {2015})},\ \Eprint
  {http://arxiv.org/abs/1506.02192} {arXiv:1506.02192} \BibitemShut {NoStop}%
\bibitem [{\citenamefont {{Hinshaw}}\ \emph {et~al.}(2013)\citenamefont
  {{Hinshaw}}, \citenamefont {{Larson}}, \citenamefont {{Komatsu}},
  \citenamefont {{Spergel}}, \citenamefont {{Bennett}}, \citenamefont
  {{Dunkley}}, \citenamefont {{Nolta}}, \citenamefont {{Halpern}},
  \citenamefont {{Hill}}, \citenamefont {{Odegard}}, \citenamefont {{Page}},
  \citenamefont {{Smith}}, \citenamefont {{Weiland}}, \citenamefont {{Gold}},
  \citenamefont {{Jarosik}}, \citenamefont {{Kogut}}, \citenamefont {{Limon}},
  \citenamefont {{Meyer}}, \citenamefont {{Tucker}}, \citenamefont
  {{Wollack}},\ and\ \citenamefont {{Wright}}}]{WMAP9}%
  \BibitemOpen
  \bibfield  {author} {\bibinfo {author} {\bibfnamefont {G.}~\bibnamefont
  {{Hinshaw}}}, \bibinfo {author} {\bibfnamefont {D.}~\bibnamefont {{Larson}}},
  \bibinfo {author} {\bibfnamefont {E.}~\bibnamefont {{Komatsu}}}, \bibinfo
  {author} {\bibfnamefont {D.~N.}\ \bibnamefont {{Spergel}}}, \bibinfo {author}
  {\bibfnamefont {C.~L.}\ \bibnamefont {{Bennett}}}, \bibinfo {author}
  {\bibfnamefont {J.}~\bibnamefont {{Dunkley}}}, \bibinfo {author}
  {\bibfnamefont {M.~R.}\ \bibnamefont {{Nolta}}}, \bibinfo {author}
  {\bibfnamefont {M.}~\bibnamefont {{Halpern}}}, \bibinfo {author}
  {\bibfnamefont {R.~S.}\ \bibnamefont {{Hill}}}, \bibinfo {author}
  {\bibfnamefont {N.}~\bibnamefont {{Odegard}}}, \bibinfo {author}
  {\bibfnamefont {L.}~\bibnamefont {{Page}}}, \bibinfo {author} {\bibfnamefont
  {K.~M.}\ \bibnamefont {{Smith}}}, \bibinfo {author} {\bibfnamefont {J.~L.}\
  \bibnamefont {{Weiland}}}, \bibinfo {author} {\bibfnamefont {B.}~\bibnamefont
  {{Gold}}}, \bibinfo {author} {\bibfnamefont {N.}~\bibnamefont {{Jarosik}}},
  \bibinfo {author} {\bibfnamefont {A.}~\bibnamefont {{Kogut}}}, \bibinfo
  {author} {\bibfnamefont {M.}~\bibnamefont {{Limon}}}, \bibinfo {author}
  {\bibfnamefont {S.~S.}\ \bibnamefont {{Meyer}}}, \bibinfo {author}
  {\bibfnamefont {G.~S.}\ \bibnamefont {{Tucker}}}, \bibinfo {author}
  {\bibfnamefont {E.}~\bibnamefont {{Wollack}}}, \ and\ \bibinfo {author}
  {\bibfnamefont {E.~L.}\ \bibnamefont {{Wright}}},\ }\href {\doibase
  10.1088/0067-0049/208/2/19} {\bibfield  {journal} {\bibinfo  {journal}
  {\apjs}\ }\textbf {\bibinfo {volume} {208}},\ \bibinfo {eid} {19} (\bibinfo
  {year} {2013})},\ \Eprint {http://arxiv.org/abs/1212.5226} {arXiv:1212.5226}
  \BibitemShut {NoStop}%
\bibitem [{\citenamefont {{Planck Collaboration}}\ \emph
  {et~al.}(2015)\citenamefont {{Planck Collaboration}}, \citenamefont {{Ade}},
  \citenamefont {{Aghanim}}, \citenamefont {{Arnaud}}, \citenamefont
  {{Ashdown}}, \citenamefont {{Aumont}}, \citenamefont {{Baccigalupi}},
  \citenamefont {{Banday}}, \citenamefont {{Barreiro}}, \citenamefont
  {{Bartlett}},\ and\ \citenamefont {et~al.}}]{PlanckCosmo}%
  \BibitemOpen
  \bibfield  {author} {\bibinfo {author} {\bibnamefont {{Planck
  Collaboration}}}, \bibinfo {author} {\bibfnamefont {P.~A.~R.}\ \bibnamefont
  {{Ade}}}, \bibinfo {author} {\bibfnamefont {N.}~\bibnamefont {{Aghanim}}},
  \bibinfo {author} {\bibfnamefont {M.}~\bibnamefont {{Arnaud}}}, \bibinfo
  {author} {\bibfnamefont {M.}~\bibnamefont {{Ashdown}}}, \bibinfo {author}
  {\bibfnamefont {J.}~\bibnamefont {{Aumont}}}, \bibinfo {author}
  {\bibfnamefont {C.}~\bibnamefont {{Baccigalupi}}}, \bibinfo {author}
  {\bibfnamefont {A.~J.}\ \bibnamefont {{Banday}}}, \bibinfo {author}
  {\bibfnamefont {R.~B.}\ \bibnamefont {{Barreiro}}}, \bibinfo {author}
  {\bibfnamefont {J.~G.}\ \bibnamefont {{Bartlett}}}, \ and\ \bibinfo {author}
  {\bibnamefont {et~al.}},\ }\href@noop {} {\bibfield  {journal} {\bibinfo
  {journal} {ArXiv e-prints}\ } (\bibinfo {year} {2015})},\ \Eprint
  {http://arxiv.org/abs/1502.01589} {arXiv:1502.01589} \BibitemShut {NoStop}%
\bibitem [{\citenamefont {{Springel}}(2005)}]{Gadget2}%
  \BibitemOpen
  \bibfield  {author} {\bibinfo {author} {\bibfnamefont {V.}~\bibnamefont
  {{Springel}}},\ }\href {\doibase 10.1111/j.1365-2966.2005.09655.x} {\bibfield
   {journal} {\bibinfo  {journal} {\mnras}\ }\textbf {\bibinfo {volume}
  {364}},\ \bibinfo {pages} {1105} (\bibinfo {year} {2005})},\ \Eprint
  {http://arxiv.org/abs/astro-ph/0505010} {astro-ph/0505010} \BibitemShut
  {NoStop}%
\bibitem [{\citenamefont {{Sato}}\ \emph {et~al.}(2009)\citenamefont {{Sato}},
  \citenamefont {{Hamana}}, \citenamefont {{Takahashi}}, \citenamefont
  {{Takada}}, \citenamefont {{Yoshida}}, \citenamefont {{Matsubara}},\ and\
  \citenamefont {{Sugiyama}}}]{Sato12}%
  \BibitemOpen
  \bibfield  {author} {\bibinfo {author} {\bibfnamefont {M.}~\bibnamefont
  {{Sato}}}, \bibinfo {author} {\bibfnamefont {T.}~\bibnamefont {{Hamana}}},
  \bibinfo {author} {\bibfnamefont {R.}~\bibnamefont {{Takahashi}}}, \bibinfo
  {author} {\bibfnamefont {M.}~\bibnamefont {{Takada}}}, \bibinfo {author}
  {\bibfnamefont {N.}~\bibnamefont {{Yoshida}}}, \bibinfo {author}
  {\bibfnamefont {T.}~\bibnamefont {{Matsubara}}}, \ and\ \bibinfo {author}
  {\bibfnamefont {N.}~\bibnamefont {{Sugiyama}}},\ }\href {\doibase
  10.1088/0004-637X/701/2/945} {\bibfield  {journal} {\bibinfo  {journal}
  {\apj}\ }\textbf {\bibinfo {volume} {701}},\ \bibinfo {pages} {945} (\bibinfo
  {year} {2009})},\ \Eprint {http://arxiv.org/abs/0906.2237} {arXiv:0906.2237
  [astro-ph.CO]} \BibitemShut {NoStop}%
\bibitem [{\citenamefont {{Takada}}\ and\ \citenamefont {{Hu}}(2013)}]{SSC1}%
  \BibitemOpen
  \bibfield  {author} {\bibinfo {author} {\bibfnamefont {M.}~\bibnamefont
  {{Takada}}}\ and\ \bibinfo {author} {\bibfnamefont {W.}~\bibnamefont
  {{Hu}}},\ }\href {\doibase 10.1103/PhysRevD.87.123504} {\bibfield  {journal}
  {\bibinfo  {journal} {\prd}\ }\textbf {\bibinfo {volume} {87}},\ \bibinfo
  {eid} {123504} (\bibinfo {year} {2013})},\ \Eprint
  {http://arxiv.org/abs/1302.6994} {arXiv:1302.6994 [astro-ph.CO]} \BibitemShut
  {NoStop}%
\bibitem [{\citenamefont {{Mohammed}}\ \emph {et~al.}(2016)\citenamefont
  {{Mohammed}}, \citenamefont {{Seljak}},\ and\ \citenamefont {{Vlah}}}]{SSC2}%
  \BibitemOpen
  \bibfield  {author} {\bibinfo {author} {\bibfnamefont {I.}~\bibnamefont
  {{Mohammed}}}, \bibinfo {author} {\bibfnamefont {U.}~\bibnamefont
  {{Seljak}}}, \ and\ \bibinfo {author} {\bibfnamefont {Z.}~\bibnamefont
  {{Vlah}}},\ }\href@noop {} {\bibfield  {journal} {\bibinfo  {journal} {ArXiv
  e-prints}\ } (\bibinfo {year} {2016})},\ \Eprint
  {http://arxiv.org/abs/1607.00043} {arXiv:1607.00043} \BibitemShut {NoStop}%
\bibitem [{\citenamefont {{Heitmann}}\ \emph {et~al.}(2015)\citenamefont
  {{Heitmann}}, \citenamefont {{Frontiere}}, \citenamefont {{Sewell}},
  \citenamefont {{Habib}}, \citenamefont {{Pope}}, \citenamefont {{Finkel}},
  \citenamefont {{Rizzi}}, \citenamefont {{Insley}},\ and\ \citenamefont
  {{Bhattacharya}}}]{Qcontinuum}%
  \BibitemOpen
  \bibfield  {author} {\bibinfo {author} {\bibfnamefont {K.}~\bibnamefont
  {{Heitmann}}}, \bibinfo {author} {\bibfnamefont {N.}~\bibnamefont
  {{Frontiere}}}, \bibinfo {author} {\bibfnamefont {C.}~\bibnamefont
  {{Sewell}}}, \bibinfo {author} {\bibfnamefont {S.}~\bibnamefont {{Habib}}},
  \bibinfo {author} {\bibfnamefont {A.}~\bibnamefont {{Pope}}}, \bibinfo
  {author} {\bibfnamefont {H.}~\bibnamefont {{Finkel}}}, \bibinfo {author}
  {\bibfnamefont {S.}~\bibnamefont {{Rizzi}}}, \bibinfo {author} {\bibfnamefont
  {J.}~\bibnamefont {{Insley}}}, \ and\ \bibinfo {author} {\bibfnamefont
  {S.}~\bibnamefont {{Bhattacharya}}},\ }\href {\doibase
  10.1088/0067-0049/219/2/34} {\bibfield  {journal} {\bibinfo  {journal}
  {\apjs}\ }\textbf {\bibinfo {volume} {219}},\ \bibinfo {eid} {34} (\bibinfo
  {year} {2015})},\ \Eprint {http://arxiv.org/abs/1411.3396} {arXiv:1411.3396}
  \BibitemShut {NoStop}%
\bibitem [{\citenamefont {{Habib}}\ \emph {et~al.}(2016)\citenamefont
  {{Habib}}, \citenamefont {{Pope}}, \citenamefont {{Finkel}}, \citenamefont
  {{Frontiere}}, \citenamefont {{Heitmann}}, \citenamefont {{Daniel}},
  \citenamefont {{Fasel}}, \citenamefont {{Morozov}}, \citenamefont
  {{Zagaris}}, \citenamefont {{Peterka}}, \citenamefont {{Vishwanath}},
  \citenamefont {{Luki{\'c}}}, \citenamefont {{Sehrish}},\ and\ \citenamefont
  {{Liao}}}]{HACC}%
  \BibitemOpen
  \bibfield  {author} {\bibinfo {author} {\bibfnamefont {S.}~\bibnamefont
  {{Habib}}}, \bibinfo {author} {\bibfnamefont {A.}~\bibnamefont {{Pope}}},
  \bibinfo {author} {\bibfnamefont {H.}~\bibnamefont {{Finkel}}}, \bibinfo
  {author} {\bibfnamefont {N.}~\bibnamefont {{Frontiere}}}, \bibinfo {author}
  {\bibfnamefont {K.}~\bibnamefont {{Heitmann}}}, \bibinfo {author}
  {\bibfnamefont {D.}~\bibnamefont {{Daniel}}}, \bibinfo {author}
  {\bibfnamefont {P.}~\bibnamefont {{Fasel}}}, \bibinfo {author} {\bibfnamefont
  {V.}~\bibnamefont {{Morozov}}}, \bibinfo {author} {\bibfnamefont
  {G.}~\bibnamefont {{Zagaris}}}, \bibinfo {author} {\bibfnamefont
  {T.}~\bibnamefont {{Peterka}}}, \bibinfo {author} {\bibfnamefont
  {V.}~\bibnamefont {{Vishwanath}}}, \bibinfo {author} {\bibfnamefont
  {Z.}~\bibnamefont {{Luki{\'c}}}}, \bibinfo {author} {\bibfnamefont
  {S.}~\bibnamefont {{Sehrish}}}, \ and\ \bibinfo {author} {\bibfnamefont
  {W.-k.}\ \bibnamefont {{Liao}}},\ }\href {\doibase
  10.1016/j.newast.2015.06.003} {\ \textbf {\bibinfo {volume} {42}},\ \bibinfo
  {pages} {49} (\bibinfo {year} {2016})},\ \Eprint
  {http://arxiv.org/abs/1410.2805} {arXiv:1410.2805 [astro-ph.IM]} \BibitemShut
  {NoStop}%
\bibitem [{\citenamefont {{Jain}}\ \emph {et~al.}(2000)\citenamefont {{Jain}},
  \citenamefont {{Seljak}},\ and\ \citenamefont {{White}}}]{RayTracingJain}%
  \BibitemOpen
  \bibfield  {author} {\bibinfo {author} {\bibfnamefont {B.}~\bibnamefont
  {{Jain}}}, \bibinfo {author} {\bibfnamefont {U.}~\bibnamefont {{Seljak}}}, \
  and\ \bibinfo {author} {\bibfnamefont {S.}~\bibnamefont {{White}}},\ }\href
  {\doibase 10.1086/308384} {\bibfield  {journal} {\bibinfo  {journal} {\apj}\
  }\textbf {\bibinfo {volume} {530}},\ \bibinfo {pages} {547} (\bibinfo {year}
  {2000})},\ \Eprint {http://arxiv.org/abs/astro-ph/9901191} {astro-ph/9901191}
  \BibitemShut {NoStop}%
\bibitem [{\citenamefont {{Hilbert}}\ \emph {et~al.}(2009)\citenamefont
  {{Hilbert}}, \citenamefont {{Hartlap}}, \citenamefont {{White}},\ and\
  \citenamefont {{Schneider}}}]{RayTracingHartlap}%
  \BibitemOpen
  \bibfield  {author} {\bibinfo {author} {\bibfnamefont {S.}~\bibnamefont
  {{Hilbert}}}, \bibinfo {author} {\bibfnamefont {J.}~\bibnamefont
  {{Hartlap}}}, \bibinfo {author} {\bibfnamefont {S.~D.~M.}\ \bibnamefont
  {{White}}}, \ and\ \bibinfo {author} {\bibfnamefont {P.}~\bibnamefont
  {{Schneider}}},\ }\href {\doibase 10.1051/0004-6361/200811054} {\bibfield
  {journal} {\bibinfo  {journal} {\aap}\ }\textbf {\bibinfo {volume} {499}},\
  \bibinfo {pages} {31} (\bibinfo {year} {2009})},\ \Eprint
  {http://arxiv.org/abs/0809.5035} {arXiv:0809.5035} \BibitemShut {NoStop}%
\bibitem [{\citenamefont {{Petri}}(2016{\natexlab{a}})}]{LensTools-ASCL}%
  \BibitemOpen
  \bibfield  {author} {\bibinfo {author} {\bibfnamefont {A.}~\bibnamefont
  {{Petri}}},\ }\href@noop {} {\enquote {\bibinfo {title} {{LensTools: Weak
  Lensing computing tools}},}\ }\bibinfo {howpublished} {Astrophysics Source
  Code Library} (\bibinfo {year} {2016}{\natexlab{a}}),\ \Eprint
  {http://arxiv.org/abs/1602.009} {ascl:1602.009} \BibitemShut {NoStop}%
\bibitem [{\citenamefont {{Petri}}(2016{\natexlab{b}})}]{LensTools-paper}%
  \BibitemOpen
  \bibfield  {author} {\bibinfo {author} {\bibfnamefont {A.}~\bibnamefont
  {{Petri}}},\ }\href {\doibase 10.1016/j.ascom.2016.06.001} {\bibfield
  {journal} {\bibinfo  {journal} {Astronomy and Computing}\ }\textbf {\bibinfo
  {volume} {17}},\ \bibinfo {pages} {73} (\bibinfo {year}
  {2016}{\natexlab{b}})},\ \Eprint {http://arxiv.org/abs/1606.01903}
  {arXiv:1606.01903} \BibitemShut {NoStop}%
\bibitem [{\citenamefont {Petri}\ \emph {et~al.}(2016)\citenamefont {Petri},
  \citenamefont {Haiman},\ and\ \citenamefont {May}}]{PetriVariance}%
  \BibitemOpen
  \bibfield  {author} {\bibinfo {author} {\bibfnamefont {A.}~\bibnamefont
  {Petri}}, \bibinfo {author} {\bibfnamefont {Z.}~\bibnamefont {Haiman}}, \
  and\ \bibinfo {author} {\bibfnamefont {M.}~\bibnamefont {May}},\ }\href
  {\doibase 10.1103/PhysRevD.93.063524} {\bibfield  {journal} {\bibinfo
  {journal} {Phys. Rev. D}\ }\textbf {\bibinfo {volume} {93}},\ \bibinfo
  {pages} {063524} (\bibinfo {year} {2016})}\BibitemShut {NoStop}%
\bibitem [{\citenamefont {{Petri}}\ \emph {et~al.}(2015)\citenamefont
  {{Petri}}, \citenamefont {{Liu}}, \citenamefont {{Haiman}}, \citenamefont
  {{May}}, \citenamefont {{Hui}},\ and\ \citenamefont
  {{Kratochvil}}}]{CFHTMink}%
  \BibitemOpen
  \bibfield  {author} {\bibinfo {author} {\bibfnamefont {A.}~\bibnamefont
  {{Petri}}}, \bibinfo {author} {\bibfnamefont {J.}~\bibnamefont {{Liu}}},
  \bibinfo {author} {\bibfnamefont {Z.}~\bibnamefont {{Haiman}}}, \bibinfo
  {author} {\bibfnamefont {M.}~\bibnamefont {{May}}}, \bibinfo {author}
  {\bibfnamefont {L.}~\bibnamefont {{Hui}}}, \ and\ \bibinfo {author}
  {\bibfnamefont {J.~M.}\ \bibnamefont {{Kratochvil}}},\ }\href {\doibase
  10.1103/PhysRevD.91.103511} {\bibfield  {journal} {\bibinfo  {journal}
  {\prd}\ }\textbf {\bibinfo {volume} {91}},\ \bibinfo {eid} {103511} (\bibinfo
  {year} {2015})},\ \Eprint {http://arxiv.org/abs/1503.06214}
  {arXiv:1503.06214} \BibitemShut {NoStop}%
\bibitem [{\citenamefont {{Liu}}\ \emph {et~al.}(2015)\citenamefont {{Liu}},
  \citenamefont {{Petri}}, \citenamefont {{Haiman}}, \citenamefont {{Hui}},
  \citenamefont {{Kratochvil}},\ and\ \citenamefont {{May}}}]{CFHTPeaks}%
  \BibitemOpen
  \bibfield  {author} {\bibinfo {author} {\bibfnamefont {J.}~\bibnamefont
  {{Liu}}}, \bibinfo {author} {\bibfnamefont {A.}~\bibnamefont {{Petri}}},
  \bibinfo {author} {\bibfnamefont {Z.}~\bibnamefont {{Haiman}}}, \bibinfo
  {author} {\bibfnamefont {L.}~\bibnamefont {{Hui}}}, \bibinfo {author}
  {\bibfnamefont {J.~M.}\ \bibnamefont {{Kratochvil}}}, \ and\ \bibinfo
  {author} {\bibfnamefont {M.}~\bibnamefont {{May}}},\ }\href {\doibase
  10.1103/PhysRevD.91.063507} {\bibfield  {journal} {\bibinfo  {journal}
  {\prd}\ }\textbf {\bibinfo {volume} {91}},\ \bibinfo {eid} {063507} (\bibinfo
  {year} {2015})},\ \Eprint {http://arxiv.org/abs/1412.0757} {arXiv:1412.0757}
  \BibitemShut {NoStop}%
\bibitem [{\citenamefont {{Zhan}}(2006)}]{HuTomo}%
  \BibitemOpen
  \bibfield  {author} {\bibinfo {author} {\bibfnamefont {H.}~\bibnamefont
  {{Zhan}}},\ }\href {\doibase 10.1088/1475-7516/2006/08/008} {\bibfield
  {journal} {\bibinfo  {journal} {JCAP}\ }\textbf {\bibinfo {volume} {8}},\
  \bibinfo {eid} {008} (\bibinfo {year} {2006})},\ \Eprint
  {http://arxiv.org/abs/astro-ph/0605696} {astro-ph/0605696} \BibitemShut
  {NoStop}%
\bibitem [{\citenamefont {{LSST Science Collaboration}}\ \emph
  {et~al.}(2009)\citenamefont {{LSST Science Collaboration}}, \citenamefont
  {{Abell}}, \citenamefont {{Allison}}, \citenamefont {{Anderson}},
  \citenamefont {{Andrew}}, \citenamefont {{Angel}}, \citenamefont {{Armus}},
  \citenamefont {{Arnett}}, \citenamefont {{Asztalos}}, \citenamefont
  {{Axelrod}},\ and\ \citenamefont {et~al.}}]{LSSTSciBook}%
  \BibitemOpen
  \bibfield  {author} {\bibinfo {author} {\bibnamefont {{LSST Science
  Collaboration}}}, \bibinfo {author} {\bibfnamefont {P.~A.}\ \bibnamefont
  {{Abell}}}, \bibinfo {author} {\bibfnamefont {J.}~\bibnamefont {{Allison}}},
  \bibinfo {author} {\bibfnamefont {S.~F.}\ \bibnamefont {{Anderson}}},
  \bibinfo {author} {\bibfnamefont {J.~R.}\ \bibnamefont {{Andrew}}}, \bibinfo
  {author} {\bibfnamefont {J.~R.~P.}\ \bibnamefont {{Angel}}}, \bibinfo
  {author} {\bibfnamefont {L.}~\bibnamefont {{Armus}}}, \bibinfo {author}
  {\bibfnamefont {D.}~\bibnamefont {{Arnett}}}, \bibinfo {author}
  {\bibfnamefont {S.~J.}\ \bibnamefont {{Asztalos}}}, \bibinfo {author}
  {\bibfnamefont {T.~S.}\ \bibnamefont {{Axelrod}}}, \ and\ \bibinfo {author}
  {\bibnamefont {et~al.}},\ }\href@noop {} {\bibfield  {journal} {\bibinfo
  {journal} {ArXiv e-prints}\ } (\bibinfo {year} {2009})},\ \Eprint
  {http://arxiv.org/abs/0912.0201} {arXiv:0912.0201 [astro-ph.IM]} \BibitemShut
  {NoStop}%
\bibitem [{\citenamefont {{Casarini}}\ \emph {et~al.}(2015)\citenamefont
  {{Casarini}}, \citenamefont {{Piattella}}, \citenamefont {{Bonometto}},\ and\
  \citenamefont {{Mezzetti}}}]{NbodySample}%
  \BibitemOpen
  \bibfield  {author} {\bibinfo {author} {\bibfnamefont {L.}~\bibnamefont
  {{Casarini}}}, \bibinfo {author} {\bibfnamefont {O.~F.}\ \bibnamefont
  {{Piattella}}}, \bibinfo {author} {\bibfnamefont {S.~A.}\ \bibnamefont
  {{Bonometto}}}, \ and\ \bibinfo {author} {\bibfnamefont {M.}~\bibnamefont
  {{Mezzetti}}},\ }\href {\doibase 10.1088/0004-637X/812/1/16} {\bibfield
  {journal} {\bibinfo  {journal} {\apj}\ }\textbf {\bibinfo {volume} {812}},\
  \bibinfo {eid} {16} (\bibinfo {year} {2015})},\ \Eprint
  {http://arxiv.org/abs/1406.5374} {arXiv:1406.5374} \BibitemShut {NoStop}%
\bibitem [{\citenamefont {{Matsubara}}(2010)}]{Matsubara10}%
  \BibitemOpen
  \bibfield  {author} {\bibinfo {author} {\bibfnamefont {T.}~\bibnamefont
  {{Matsubara}}},\ }\href {\doibase 10.1103/PhysRevD.81.083505} {\bibfield
  {journal} {\bibinfo  {journal} {\prd}\ }\textbf {\bibinfo {volume} {81}},\
  \bibinfo {eid} {083505} (\bibinfo {year} {2010})},\ \Eprint
  {http://arxiv.org/abs/1001.2321} {arXiv:1001.2321 [astro-ph.CO]} \BibitemShut
  {NoStop}%
\bibitem [{\citenamefont {{Munshi}}\ \emph {et~al.}(2012)\citenamefont
  {{Munshi}}, \citenamefont {{van Waerbeke}}, \citenamefont {{Smidt}},\ and\
  \citenamefont {{Coles}}}]{Munshi12}%
  \BibitemOpen
  \bibfield  {author} {\bibinfo {author} {\bibfnamefont {D.}~\bibnamefont
  {{Munshi}}}, \bibinfo {author} {\bibfnamefont {L.}~\bibnamefont {{van
  Waerbeke}}}, \bibinfo {author} {\bibfnamefont {J.}~\bibnamefont {{Smidt}}}, \
  and\ \bibinfo {author} {\bibfnamefont {P.}~\bibnamefont {{Coles}}},\ }\href
  {\doibase 10.1111/j.1365-2966.2011.19718.x} {\bibfield  {journal} {\bibinfo
  {journal} {\mnras}\ }\textbf {\bibinfo {volume} {419}},\ \bibinfo {pages}
  {536} (\bibinfo {year} {2012})},\ \Eprint {http://arxiv.org/abs/1103.1876}
  {arXiv:1103.1876 [astro-ph.CO]} \BibitemShut {NoStop}%
\bibitem [{\citenamefont {{Petri}}\ \emph {et~al.}(2013)\citenamefont
  {{Petri}}, \citenamefont {{Haiman}}, \citenamefont {{Hui}}, \citenamefont
  {{May}},\ and\ \citenamefont {{Kratochvil}}}]{MinkPetri}%
  \BibitemOpen
  \bibfield  {author} {\bibinfo {author} {\bibfnamefont {A.}~\bibnamefont
  {{Petri}}}, \bibinfo {author} {\bibfnamefont {Z.}~\bibnamefont {{Haiman}}},
  \bibinfo {author} {\bibfnamefont {L.}~\bibnamefont {{Hui}}}, \bibinfo
  {author} {\bibfnamefont {M.}~\bibnamefont {{May}}}, \ and\ \bibinfo {author}
  {\bibfnamefont {J.~M.}\ \bibnamefont {{Kratochvil}}},\ }\href {\doibase
  10.1103/PhysRevD.88.123002} {\bibfield  {journal} {\bibinfo  {journal}
  {\prd}\ }\textbf {\bibinfo {volume} {88}},\ \bibinfo {eid} {123002} (\bibinfo
  {year} {2013})},\ \Eprint {http://arxiv.org/abs/1309.4460} {arXiv:1309.4460}
  \BibitemShut {NoStop}%
\bibitem [{\citenamefont {{Ivezi{\'c}}}\ \emph {et~al.}(2014)\citenamefont
  {{Ivezi{\'c}}}, \citenamefont {{Connolly}}, \citenamefont {{Vanderplas}},\
  and\ \citenamefont {{Gray}}}]{astroMLText}%
  \BibitemOpen
  \bibfield  {author} {\bibinfo {author} {\bibfnamefont {{\v Z}.}~\bibnamefont
  {{Ivezi{\'c}}}}, \bibinfo {author} {\bibfnamefont {A.}~\bibnamefont
  {{Connolly}}}, \bibinfo {author} {\bibfnamefont {J.}~\bibnamefont
  {{Vanderplas}}}, \ and\ \bibinfo {author} {\bibfnamefont {A.}~\bibnamefont
  {{Gray}}},\ }\href@noop {} {\emph {\bibinfo {title} {Statistics, Data Mining
  and Machine Learning in Astronomy}}}\ (\bibinfo  {publisher} {Princeton
  University Press},\ \bibinfo {year} {2014})\BibitemShut {NoStop}%
\bibitem [{\citenamefont {{Hartlap}}\ \emph {et~al.}(2007)\citenamefont
  {{Hartlap}}, \citenamefont {{Simon}},\ and\ \citenamefont
  {{Schneider}}}]{Hartlap07}%
  \BibitemOpen
  \bibfield  {author} {\bibinfo {author} {\bibfnamefont {J.}~\bibnamefont
  {{Hartlap}}}, \bibinfo {author} {\bibfnamefont {P.}~\bibnamefont {{Simon}}},
  \ and\ \bibinfo {author} {\bibfnamefont {P.}~\bibnamefont {{Schneider}}},\
  }\href {\doibase 10.1051/0004-6361:20066170} {\bibfield  {journal} {\bibinfo
  {journal} {\aap}\ }\textbf {\bibinfo {volume} {464}},\ \bibinfo {pages} {399}
  (\bibinfo {year} {2007})},\ \Eprint
  {http://arxiv.org/abs/arXiv:astro-ph/0608064} {arXiv:astro-ph/0608064}
  \BibitemShut {NoStop}%
\bibitem [{\citenamefont {{Taylor}}\ \emph {et~al.}(2013)\citenamefont
  {{Taylor}}, \citenamefont {{Joachimi}},\ and\ \citenamefont
  {{Kitching}}}]{Taylor12}%
  \BibitemOpen
  \bibfield  {author} {\bibinfo {author} {\bibfnamefont {A.}~\bibnamefont
  {{Taylor}}}, \bibinfo {author} {\bibfnamefont {B.}~\bibnamefont
  {{Joachimi}}}, \ and\ \bibinfo {author} {\bibfnamefont {T.}~\bibnamefont
  {{Kitching}}},\ }\href {\doibase 10.1093/mnras/stt270} {\bibfield  {journal}
  {\bibinfo  {journal} {\mnras}\ }\textbf {\bibinfo {volume} {432}},\ \bibinfo
  {pages} {1928} (\bibinfo {year} {2013})},\ \Eprint
  {http://arxiv.org/abs/1212.4359} {arXiv:1212.4359} \BibitemShut {NoStop}%
\bibitem [{\citenamefont {{Taylor}}\ and\ \citenamefont
  {{Joachimi}}(2014)}]{Taylor14}%
  \BibitemOpen
  \bibfield  {author} {\bibinfo {author} {\bibfnamefont {A.}~\bibnamefont
  {{Taylor}}}\ and\ \bibinfo {author} {\bibfnamefont {B.}~\bibnamefont
  {{Joachimi}}},\ }\href {\doibase 10.1093/mnras/stu996} {\bibfield  {journal}
  {\bibinfo  {journal} {\mnras}\ }\textbf {\bibinfo {volume} {442}},\ \bibinfo
  {pages} {2728} (\bibinfo {year} {2014})},\ \Eprint
  {http://arxiv.org/abs/1402.6983} {arXiv:1402.6983} \BibitemShut {NoStop}%
\bibitem [{Note1()}]{Note1}%
  \BibitemOpen
  \bibinfo {note} {The archive we used is located \protect \url
  {http://pla.esac.esa.int/pla/}; we used the MCMC chains contained in the
  \protect \texttt {base\protect \_w/plikHM\protect \_TT\protect \_lowTEB}
  directory, labeled as \protect \texttt {base\protect \_w\protect
  \_plikHM\protect \_TT\protect \_lowTEB\protect \_[1-4].txt}}\BibitemShut
  {NoStop}%
\bibitem [{\citenamefont {{Kilbinger}}\ \emph {et~al.}(2009)\citenamefont
  {{Kilbinger}}, \citenamefont {{Benabed}}, \citenamefont {{Guy}},
  \citenamefont {{Astier}}, \citenamefont {{Tereno}}, \citenamefont {{Fu}},
  \citenamefont {{Wraith}}, \citenamefont {{Coupon}}, \citenamefont
  {{Mellier}}, \citenamefont {{Balland}}, \citenamefont {{Bouchet}},
  \citenamefont {{Hamana}}, \citenamefont {{Hardin}}, \citenamefont
  {{McCracken}}, \citenamefont {{Pain}}, \citenamefont {{Regnault}},
  \citenamefont {{Schultheis}},\ and\ \citenamefont {{Yahagi}}}]{Nicaea}%
  \BibitemOpen
  \bibfield  {author} {\bibinfo {author} {\bibfnamefont {M.}~\bibnamefont
  {{Kilbinger}}}, \bibinfo {author} {\bibfnamefont {K.}~\bibnamefont
  {{Benabed}}}, \bibinfo {author} {\bibfnamefont {J.}~\bibnamefont {{Guy}}},
  \bibinfo {author} {\bibfnamefont {P.}~\bibnamefont {{Astier}}}, \bibinfo
  {author} {\bibfnamefont {I.}~\bibnamefont {{Tereno}}}, \bibinfo {author}
  {\bibfnamefont {L.}~\bibnamefont {{Fu}}}, \bibinfo {author} {\bibfnamefont
  {D.}~\bibnamefont {{Wraith}}}, \bibinfo {author} {\bibfnamefont
  {J.}~\bibnamefont {{Coupon}}}, \bibinfo {author} {\bibfnamefont
  {Y.}~\bibnamefont {{Mellier}}}, \bibinfo {author} {\bibfnamefont
  {C.}~\bibnamefont {{Balland}}}, \bibinfo {author} {\bibfnamefont {F.~R.}\
  \bibnamefont {{Bouchet}}}, \bibinfo {author} {\bibfnamefont {T.}~\bibnamefont
  {{Hamana}}}, \bibinfo {author} {\bibfnamefont {D.}~\bibnamefont {{Hardin}}},
  \bibinfo {author} {\bibfnamefont {H.~J.}\ \bibnamefont {{McCracken}}},
  \bibinfo {author} {\bibfnamefont {R.}~\bibnamefont {{Pain}}}, \bibinfo
  {author} {\bibfnamefont {N.}~\bibnamefont {{Regnault}}}, \bibinfo {author}
  {\bibfnamefont {M.}~\bibnamefont {{Schultheis}}}, \ and\ \bibinfo {author}
  {\bibfnamefont {H.}~\bibnamefont {{Yahagi}}},\ }\href {\doibase
  10.1051/0004-6361/200811247} {\bibfield  {journal} {\bibinfo  {journal}
  {\aap}\ }\textbf {\bibinfo {volume} {497}},\ \bibinfo {pages} {677} (\bibinfo
  {year} {2009})},\ \Eprint {http://arxiv.org/abs/0810.5129} {arXiv:0810.5129}
  \BibitemShut {NoStop}%
\bibitem [{\citenamefont {{Dodelson}}\ and\ \citenamefont
  {{Schneider}}(2013)}]{DodelsonSchneider13}%
  \BibitemOpen
  \bibfield  {author} {\bibinfo {author} {\bibfnamefont {S.}~\bibnamefont
  {{Dodelson}}}\ and\ \bibinfo {author} {\bibfnamefont {M.~D.}\ \bibnamefont
  {{Schneider}}},\ }\href {\doibase 10.1103/PhysRevD.88.063537} {\bibfield
  {journal} {\bibinfo  {journal} {\prd}\ }\textbf {\bibinfo {volume} {88}},\
  \bibinfo {eid} {063537} (\bibinfo {year} {2013})},\ \Eprint
  {http://arxiv.org/abs/1304.2593} {arXiv:1304.2593 [astro-ph.CO]} \BibitemShut
  {NoStop}%
\end{thebibliography}%

\label{lastpage}
\end{document}